\documentclass[9pt,twocolumn,twoside]{osajnl}

\journal{pr}

\usepackage[utf8]{inputenc}
\usepackage[T1]{fontenc}

\usepackage[english]{babel}

\usepackage{amsmath,amssymb,amsfonts}
\usepackage{amstext, mathrsfs, textcomp}
\usepackage{mathptmx}
\usepackage{bigints}  
\usepackage{nicefrac}
\usepackage{multirow}
\usepackage{multicol}
\usepackage{dcolumn}
\usepackage{bm}

\usepackage{subfigure}
\usepackage{graphicx}
\usepackage{xcolor}
\usepackage[export]{adjustbox}

\usepackage[final]{changes}

\usepackage[breakable, most]{tcolorbox}
\tcbset{colback=blue!5!white,colframe=blue!75!black,fonttitle=\bfseries}

\graphicspath{{Figures/}{TOC_ACS/}}

\newcommand{\TITLE}{Deep learning in nano-photonics: inverse design and beyond}

\title{\TITLE}
\author[1,*]{Peter R. Wiecha}
\author[2,*]{Arnaud Arbouet}
\author[2,*]{Christian Girard}
\author[3,*]{Otto L. Muskens}

\affil[1]{LAAS, Universit\'e de Toulouse, CNRS, Toulouse, France}
\affil[2]{CEMES, Université de Toulouse, CNRS, Toulouse, France}
\affil[3]{Physics and Astronomy, Faculty of Engineering and Physical Sciences, University of Southampton, Southampton, UK}
\affil[*]{pwiecha@laas.fr}
\affil[*]{arbouet@cemes.fr}
\affil[*]{girard@cemes.fr}
\affil[*]{o.muskens@soton.ac.uk}

\doi{\url{http://dx.doi.org/XX.XXXX/XX.XX.XXXXXX}}

\begin{abstract}
	Deep learning in the context of nano-photonics is mostly discussed in terms of its potential for inverse design of photonic devices or nanostructures. Many of the recent works on machine-learning inverse design are highly specific, and the drawbacks of the respective approaches are often not immediately clear. In this review we want therefore to provide a critical review on the capabilities of deep learning for inverse design and the progress which has been made so far. We classify the different deep learning-based inverse design approaches at a higher level as well as by the context of their respective applications and critically discuss their strengths and weaknesses. While a significant part of the community’s attention lies on nano-photonic inverse design, deep learning has evolved as a tool for a large variety of applications. The second part of the review will focus therefore on machine learning research in nano-photonics ``beyond inverse design''. This spans from physics informed neural networks for tremendous acceleration of photonics simulations, over sparse data reconstruction, imaging and ``knowledge discovery'' to experimental applications.
\end{abstract}

\setboolean{displaycopyright}{false}

\newcounter{BoxCounter}

\begin{document}
	
	\maketitle

	
	\tableofcontents

\section{Introduction}

Light-matter interaction at sub-wavelength dimensions can lead to astonishing effects like localized surface plasmon resonances which concentrate light to deeply sub-wavelength volumes \cite{muhlschlegelResonantOpticalAntennas2005}, the appearance of optical magnetic resonances in otherwise non-magnetic media \cite{kuznetsovOpticallyResonantDielectric2016}, the possibility to shape optical near-fields with sub-wavelength structure \cite{girardFieldsNanostructures2005}, the emergence of non-linear optical phenomena \cite{kauranenNonlinearPlasmonics2012} or strong enhancement of quantum emitter luminescence \cite{francsPlasmonicPurcellFactor2016} to name just a few.
Those nano-scale optical effects can be exploited for a broad variety of applications, for instance in integrated quantum optics \cite{wangIntegratedPhotonicQuantum2020}, for meta-materials \cite{pendryNegativeRefractionMakes2000} and in this context specifically for metasurfaces like flat lenses \cite{genevetRecentAdvancesPlanar2017}. 
It is for example even possible to create all-optical devices which use light to solve integral equations or perform other analogue optical computing tasks \cite{estakhriInversedesignedMetastructuresThat2019, clementsOptimalDesignUniversal2016, zangeneh-nejadAnalogueComputingMetamaterials2020}.

\refstepcounter{BoxCounter}\label{box:ann_introduction}
\begin{tcolorbox}[breakable,float*=t, width=\textwidth, title={\textbf{Box \arabic{BoxCounter}.} Artificial neurons, neural networks and their training}]
	\begin{multicols}{2}
		
		An artificial neuron (AN) is simply a mathematical function which mimics the behavior of a biological neuron. 
		
		\begin{center}
			\includegraphics[width=0.95\columnwidth]{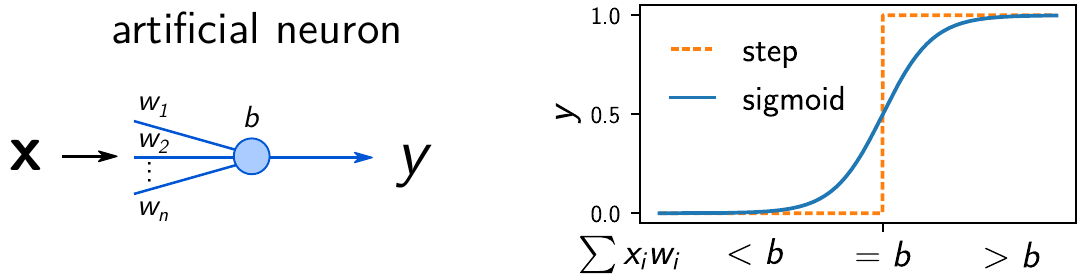}
		\end{center}
		
		The step-like behavior of neuronal activation, which starts to fire once a threshold stimulation is exceeded, can be implemented by various nonlinear mathematical functions. 
		A popular example is the logistic function (also called ``sigmoid''), shown in the above sketch.
		If the scalar product of an input vector $\mathbf{x}$ and the neuron-intrinsic weight parameters $w_i$ is larger than the neuron's bias parameter $b$, the output $y$ is ``high'' (the artificial neuron fires). Else it is ``low''. 
		
		An artificial neural network (ANN) is composed of several of such ANs, usually arranged in ``layers''. 
		The output value of a neuron is fed into a succeeding layer of neurons. The final layer is the network output $\mathbf{y}$. 
		For instance in a so-called \textit{fully connected} ANN, every neuron of one layer is connected to every neuron of the following layer:
		
		\begin{center}
			\includegraphics[width=0.95\columnwidth]{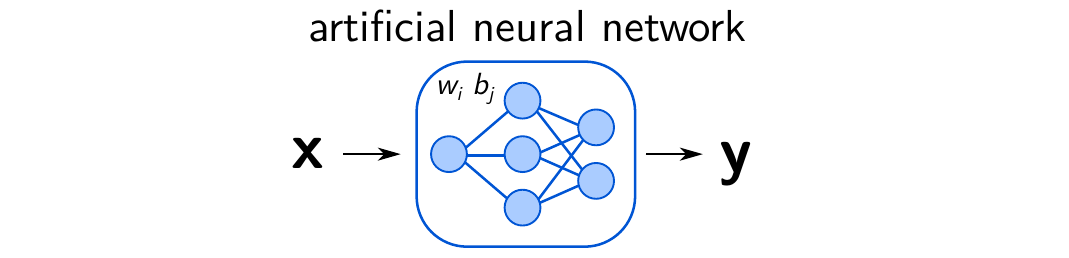}
		\end{center}
		
		Hence, an ANN represents a vectorial function $f(\mathbf{x}) = \mathbf{y}$ characterized by a large number of parameters $w_i$ and $b_j$.

		Training an ANN is done via a numerical minimization of a loss function $L$, which describes the error of the network in predicting samples of the training data.
		A popular loss function is the mean square error loss (MSE), in particular used for regression tasks:
		\begin{equation}
			L(w_i, b_j) = \frac{1}{N} \sum\limits_{l=1}^N \big( \mathbf{y}_{\text{train},l} - \mathbf{y}_{\text{ANN}}(\mathbf{x}_{\text{train},l}) \big)^2
		\end{equation}
		\(\mathbf{x}_{\text{train}}\) and \(\mathbf{y}_{\text{train}}\) are $N$ random samples from the training data, \(\mathbf{y}_{\text{ANN}}\) the network predictions corresponding to \(\mathbf{x}_{\text{train}}\). $N$ is called the \emph{batch size}.
		
		The term ``learning'' refers to optimizing the parameters $w_i$ and $b_j$ describing the ANN, with the goal to minimize the loss $L$.
		A small loss means that the network output approximates well the training data, ideally by learning to ``understand'' the underlying correlations.
		$L$ is numerically minimized by ``slipping down'' its gradient with respect to the parameters $w_i$ and $b_j$.
		
		\begin{center}
			\includegraphics[width=0.8\columnwidth]{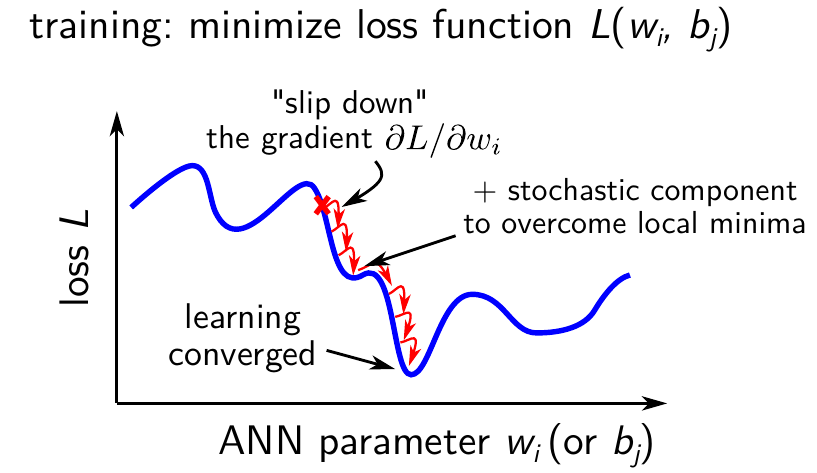}
		\end{center}
		
		Training on small batches composed of random subsets of $N$ training samples helps to ``jump'' out of local minima by adding a stochastic component to the procedure. 
		One of the most common training algorithms is \emph{Stochastic Gradient Descent} (``SGD'') \cite{goodfellowDeepLearning2016}.
		
	\end{multicols}
\end{tcolorbox}

Still, ever since the advent of nano-optics with the invention of near-field microscopy \cite{ashSuperresolutionApertureScanning1972, pohlOpticalStethoscopyImage1984, betzigNearfieldDiffractionSlit1986} the numerical description of many problems continues to be challenging \cite{gallinetNumericalMethodsNanophotonics2015}.
An example is the rational design of nano-photonic structures for specific tasks which remains a general problem that often involves brute force ``forward'' calculations, or solving inverse scattering problems.
Other challenges in nano-optics are related to experimental limitations such as the stochastic nature of single-photon emitters, fluctuating nanoscale force fields such as Brownian motion, and the diffraction limit blocking access to sub-wavelength information. 
Such effects often complicate the interpretation of nano-optics experiments and require the use of more sophisticated techniques for data analysis, for example combining data with prior knowledge or sparsity constraints.
All these obstacles are about to be pushed significantly further by the emerging computational methods around machine learning.
Especially ``deep learning'', a sub-field of machine learning which uses complex artificial neural networks (ANNs) with millions of artificial neurons (ANs), recently emerged as versatile and powerful numerical tool \cite{lecunDeepLearning2015, goodfellowDeepLearning2016}.
Deep learning techniques have proven to be particularly good at the categorization of huge and complex datasets, a task that they perform radically differently compared to classical algorithms. 
Following a rather ``intuitive'' approach, ANNs mimic the working principle of biological neurons and the human brain.
A brief overview of the basic concepts is given in Box~\ref{box:ann_introduction}.

Research in medicine is often of statistical nature, for which data-driven analysis methods like deep learning are particularly interesting.
Consequently, one of the first scientific fields to which deep learning methods have been extensively applied was medical research.
In medical diagnostics, especially medical imaging such as radiology, the use of machine learning techniques for analysis and interpretation has literally exploded in the recent past, which lead to extraordinary successes with diagnostic classification accuracies often far beyond human performance \cite{chanWillMachineLearning2018, lundervoldOverviewDeepLearning2019}.

In nano-optics and photonics, machine learning started to emerge a little later, but recently celebrated some remarkable breakthroughs, enabling the analysis, categorization and interpretation of data which seemed formerly impossible.
While already back in the 1990s simple ANNs had been discussed and used for applications in spectroscopy or for automated instrumental control for instance to counteract drifts in microscopy \cite{cirovicFeedforwardArtificialNeural1997}, it took two decades before the available computational power reached a level, that deep ANNs with millions or even up to hundreds of billions of free parameters \cite{brownLanguageModelsAre2020} could be successfully trained on formerly unsolved problems. 
Today, deep learning models have evolved to an extent that they readily outperform humans on specialized tasks like image recognition \cite{lecunDeepLearning2015, szegedyInceptionv4InceptionResNetImpact2016}.
This progress was possible especially thanks to the rapid development of massively parallel computing architectures in modern GPUs, and lately of specific ``tensor cores'', integrated logic circuits optimized for the mathematical matrix operation tasks required for neural network training.
Even all-optical implementations of artificial neural networks have been subject to recent research, however their performance is still limited by the lack of energy-efficient all-optical non-linear units \cite{linAllopticalMachineLearning2018, hughesWavePhysicsAnalog2019, menguScaleShiftRotationinvariant2020}.

%
%
%
%
%
%
%
\begin{figure*}[t!]
	\centering{
		\includegraphics[width=0.95\linewidth]{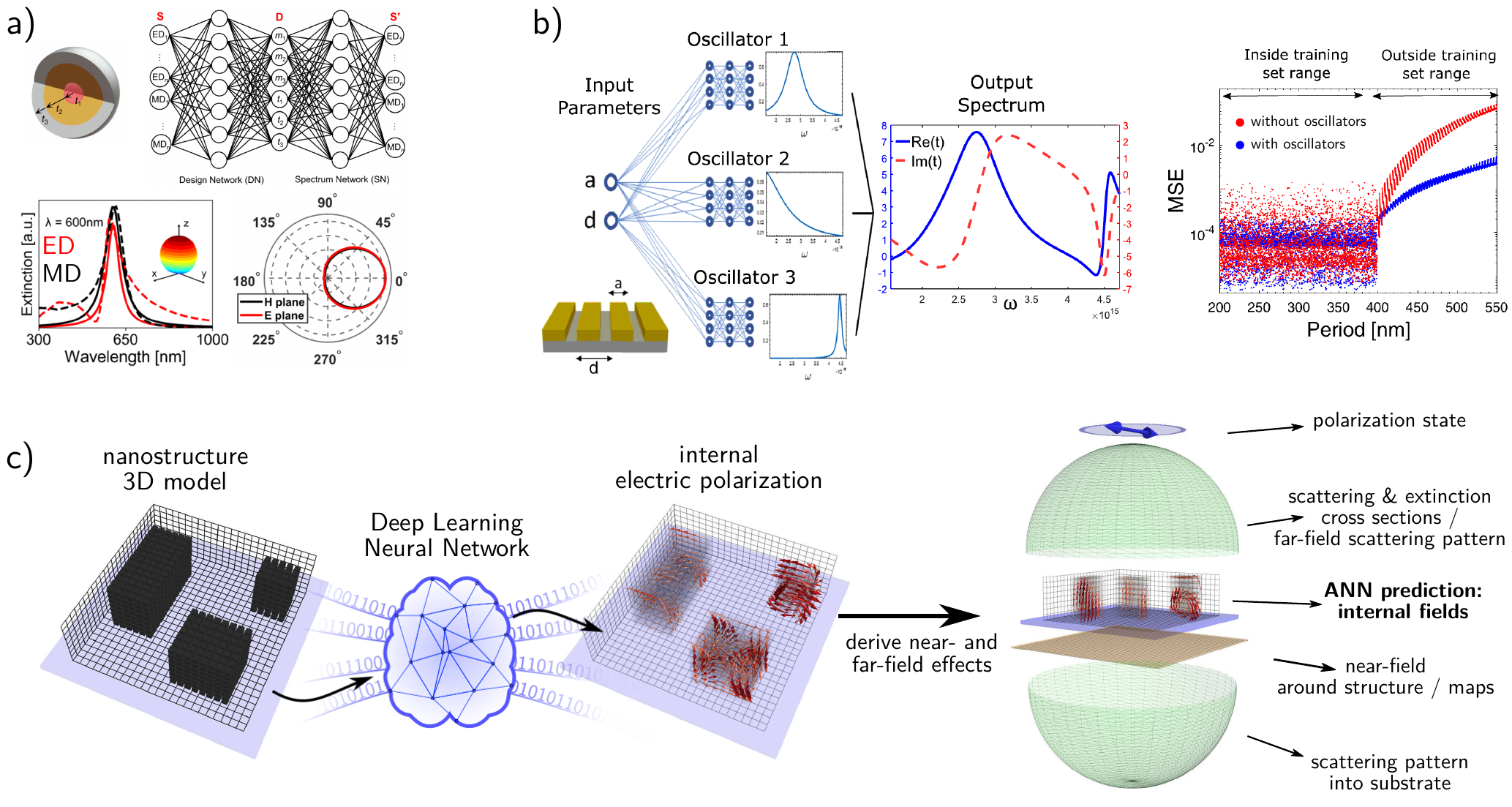}}
	\caption{
		Deep learning based forward solvers for ultra-fast physics predictions.
		(a) simultaneous electric and magnetic dipole resonance prediction and inverse design in multi-layer nano-spheres. Adapted with permission from \cite{soSimultaneousInverseDesign2019}. Copyright (2019) American Chemical Society. 
		(b) nano-optics solver network, which predicts the optical response of a grating based on multiple Lorentz-oscillators. As shown in the right panel, the physics based data representation allows the network to generalize well outside the range of the training data (blue points). Adapted with permission from \cite{blanchard-dionneTeachingOpticsMachine2020}. Copyright OSA, 2020.
		(c) internal electric polarization density predictor network. The results can be used in a coupled dipole approximation framework to calculate a large number of secondary near- and far-field effects. Adapted with permission from \cite{wiechaDeepLearningMeets2020}. Copyright (2020) American Chemical Society.
	}
	\label{fig:forward_models}
\end{figure*}

\subsection*{This work and its positioning with respect to other reviews}

Several review articles have been published recently, which categorize in great detail the latest developments of deep learning applications in photonics and nano-optics. 
For an exhaustive overview we therefore invite the reader to consult these articles \cite{hegdeDeepLearningNew2020, maDeepLearningDesign2020, soDeepLearningEnabled2020, jiangDeepNeuralNetworks2020, huangDeepLearningEnabled2020}. 
Also a few thematically more distantly related review articles have been published recently, which we want to indicate to the interested reader. 
They cover for example conventional inverse design and optimization methods for metasurfaces \cite{elsawyNumericalOptimizationMethods2020} and nano-photonics \cite{moleskyInverseDesignNanophotonics2018}, but also a few more general reviews on artificial intelligence in nanotechnology, photonics and for light-matter interaction have been published  \cite{sachaArtificialIntelligenceNanotechnology2013, yaoIntelligentNanophotonicsMerging2019, zhouEmergingRoleMachine2019, piccinottiArtificialIntelligencePhotonics2020}.
Finally, for the sake of conciseness of this review, we intentionally ignore the vast and very active research field on hardware implementations of artificial neural networks which includes -- but is not limited to -- research efforts on photonics platforms \cite{moughamesThreedimensionalWaveguideInterconnects2020, porteCompleteParallelAutonomous2020, linAllopticalMachineLearning2018}.

In this mini-review we focus on selected key results that have recently led to breakthrough advancements in the research on inverse design of photonic nanostructures and metasurfaces.
Rather than compiling an exhaustive catalogue of every single publication, we provide an overview of milestone concepts for improving deep learning inverse design fidelity, which recently allowed to bring ANNs closer to the performance of conventional optimization methods. 
We believe that such a summary of concepts is of particular interest for researchers in the field.
We dedicate the second part of the review to an overview of original applications of deep learning in nano-photonics beyond structural inverse design. 
Specifically we summarize recent developments around physics informed neural networks in optics, on deep learning for knowledge discovery and explainable machine learning as well as on applications of ANNs to nano-photonics experiments.

\section{Deep learning based nano-photonics inverse design}

The first part of this mini-review is dedicated to deep learning based inverse design techniques as well as to concepts to improve the inverse design model fidelity.
As stated before, we do not aim to provide an exhaustive list of applications. 
An up-to-date and very complete overview of possible optimization targets can be found for instance in the recent reviews by Ma et al. \cite{maDeepLearningDesign2020} or by Jiang et al. \cite{jiangDeepNeuralNetworks2020}.

\subsection*{``Conventional'' inverse design methods}

Before the recent rise of deep learning methods, inverse design of nano-photonic structures was often based on intuitive considerations and systematic fine-tuning (see e.g. \cite{blackOptimalPolarizationConversion2014, celebranoModeMatchingMultiresonant2015}).
A more systematic alternative was the combination of numerical simulation methods with gradient based or heuristic optimization algorithms like stimulated annealing, topology optimization or genetic algorithms \cite{jensenTopologyOptimizationNanophotonics2011, moleskyInverseDesignNanophotonics2018, campbellReviewNumericalOptimization2019, mengBidirectionalEvolutionaryOptimization2015, osherFrontsPropagatingCurvaturedependent1988}.
Such methods led to some remarkable success for instance in the optimization of plasmonic optical antennas \cite{feichtnerEvolutionaryOptimizationOptical2012, wiechaDesignPlasmonicDirectional2019}, dielectric multi-functional nanostructures \cite{wiechaEvolutionaryMultiobjectiveOptimization2017} or metasurfaces \cite{zhuOptimalHighEfficiency2019, elsawyNumericalOptimizationMethods2020}.
A great advantage of such methods is the possibility to include fabrication constraints or robustness conditions in the optimization procedure \cite{augensteinInverseDesignNanophotonic2020, wiechaEvolutionaryMultiobjectiveOptimization2017}.

However, heuristics coupled to numerical simulation techniques is slow and computationally expensive. Furthermore, for each new optimization target, the parameter-space needs to be searched from scratch, implying hundreds to thousands of numerical simulations. 
The recent advent of data-driven techniques like deep learning holds promise to accelerate the computation by many orders of magnitude and quite some remarkable progress has been made in the past few years.
One can distinguish two types of approach that have gained traction. The first one replaces the forward simulation in an iterative optimization with an ANN, while the second aims to build an inverse ANN that solves the problem directly. Below we critically discuss the two approaches as well as efforts at improving the quality of results.

\begin{figure*}[t!]
	\centering{
		\includegraphics[width=1.0\linewidth]{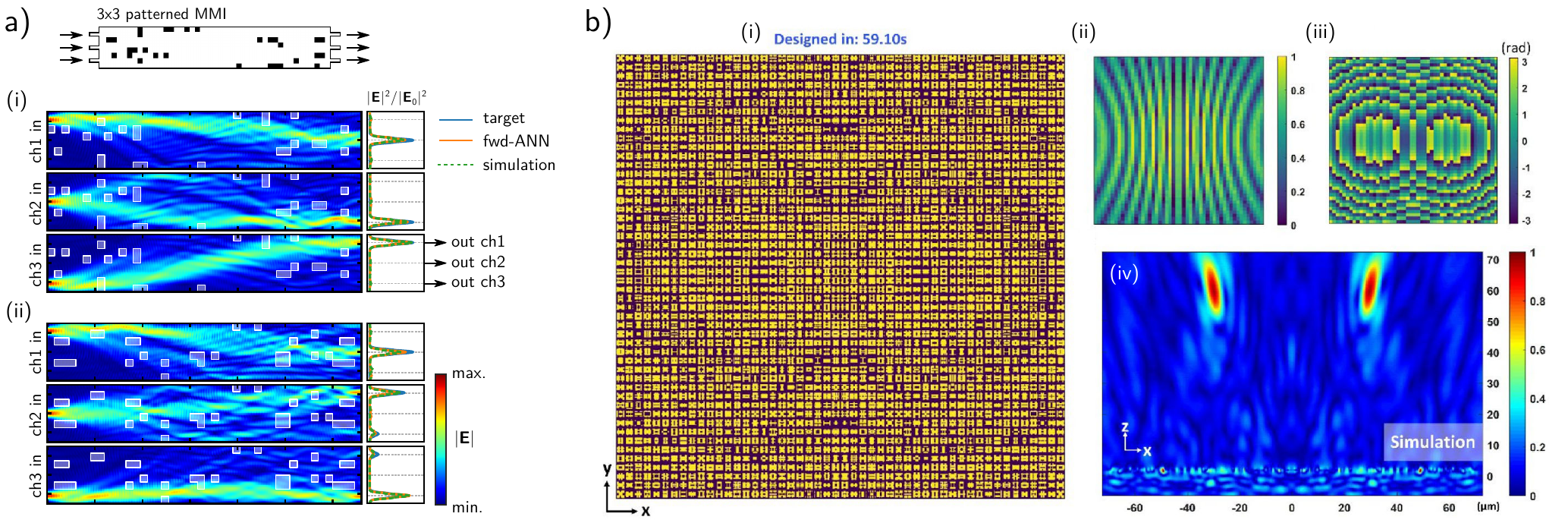}}
	\caption{
		Examples of devices inverse designed by ML algorithms. 
		a) encoder-decoder type tandem inverse network used to design perturbation patterns for $3\times 3$ MMIs as arbitrary transmission matrix elements. The light routing behavior of the second and the third input channels is interchanged between cases (i) and (ii), whilst the first input channel keeps routing light to the second output. Adapted with permission from \cite{dinsdaleDeepLearningEnabled2021}. Copyright (2021) American Chemical Society.
		b) double-focus flat lens designed by a conditional WGAN inverse network. (i) shows the dielectric metasurface, (ii) the corresponding amplitude and (iii) the phase mask. (iv) shows a numerical simulation of the field intensity to test the ANN-design. Adapted with permission from \cite{anMultifunctionalMetasurfaceDesign2020}.
	}
	\label{fig:ML_inverse_design}
\end{figure*}

\subsection{Surrogate model based inverse design}

Deep learning models are particularly strong in predicting approximate solutions to direct problems like the optical response of photonic structures. 
A possible approach to accelerate inverse design is therefore to use a ``forward neural network'' as an ultra-fast predictor together with an optimization technique.
In such a case the ANN acts as a so-called \textit{surrogate model}, taking the place of the much slower conventional simulation method.

\refstepcounter{BoxCounter}\label{box:one_to_many}
\begin{tcolorbox}[breakable,float*=t, width=\textwidth, title=\textbf{Box \arabic{BoxCounter}.} Inverse design: The one-to-many problem]
	\begin{multicols}{2}
		
		Let's assume a simple toy problem: 
		Under fixed wavelength illumination, we want to tailor the extinction coefficient of a gold nanorod by varying its length:
		
		\begin{center}
			\includegraphics[width=0.65\columnwidth]{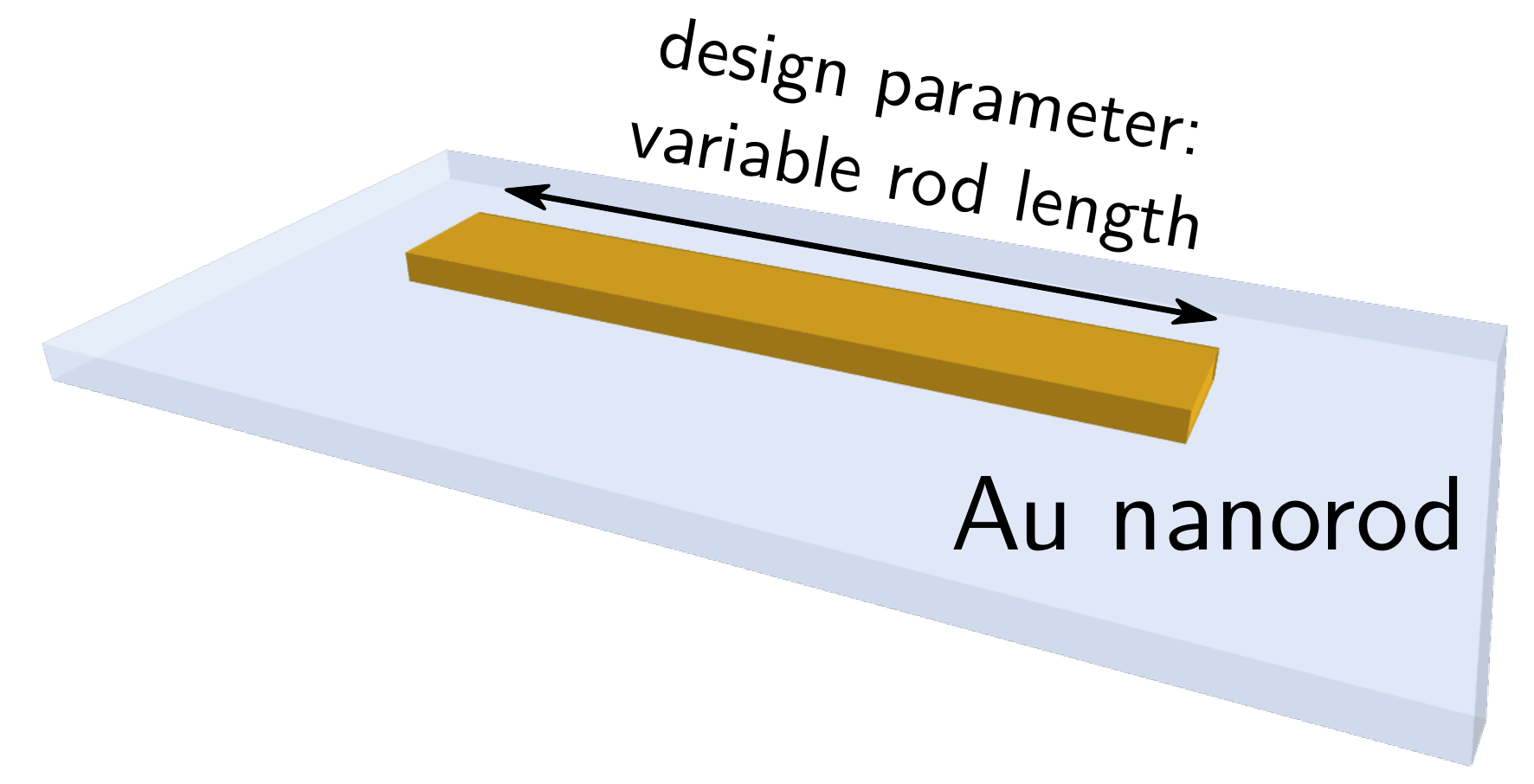}
		\end{center}
		
		Already this simple problem is ambiguous: Several rod lengths can lead to the same extinction, which makes a naive ANN implementation fail in those cases.
		
		\begin{center}
			\includegraphics[width=0.95\columnwidth]{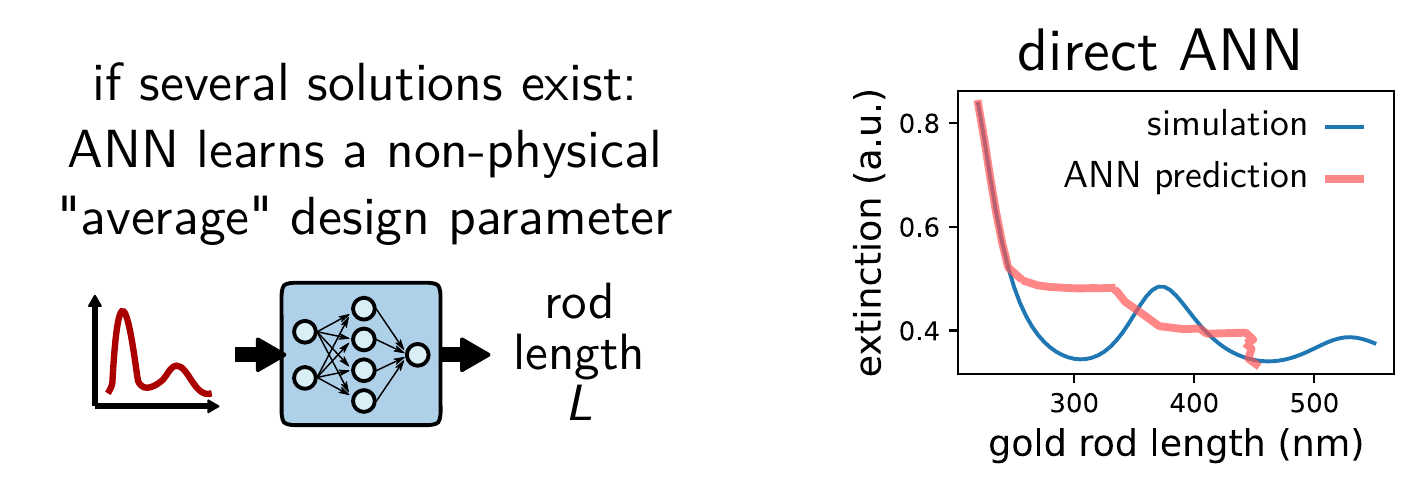}
		\end{center}

		The ``tandem neural network'' can stabilize the generator (G) via a physics loss, based on a pre-trained forward model (fwd). This approach however limits the inverse design to one solution per design target, rendering inaccessible possible multiple solutions to a given problem:
		\begin{center}
			\includegraphics[width=0.95\columnwidth]{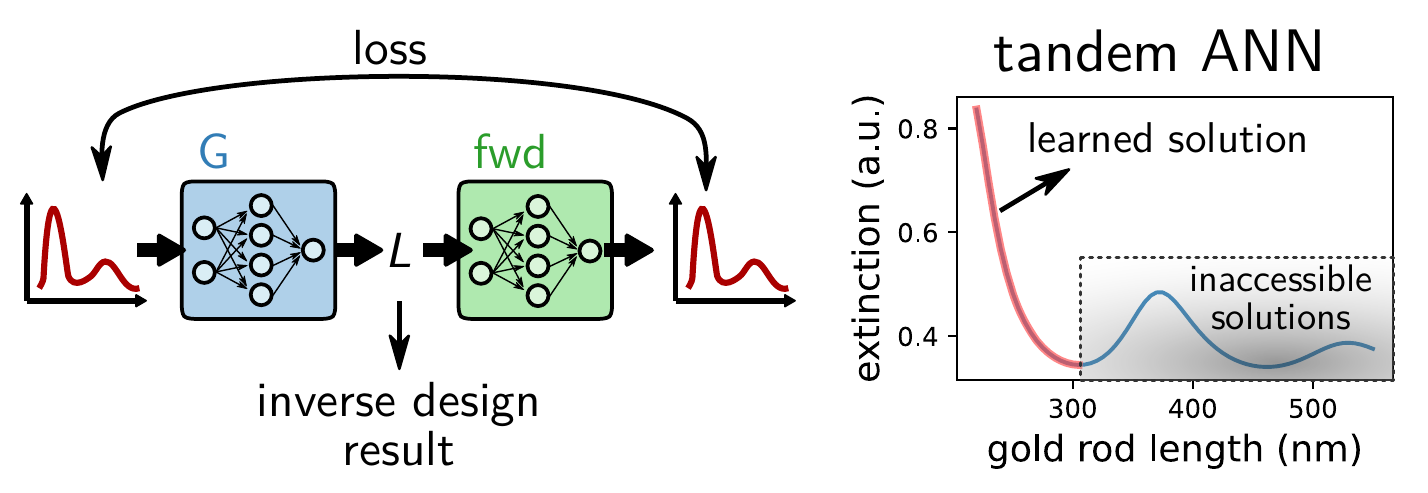}
		\end{center}
		
		Mixture density networks predict several possible solutions at a time including their respective importance as Gaussian distributions. A disadvantage is that the maximum number of possible simultaneous solutions needs to be known.
		\begin{center}
			\includegraphics[width=0.95\columnwidth]{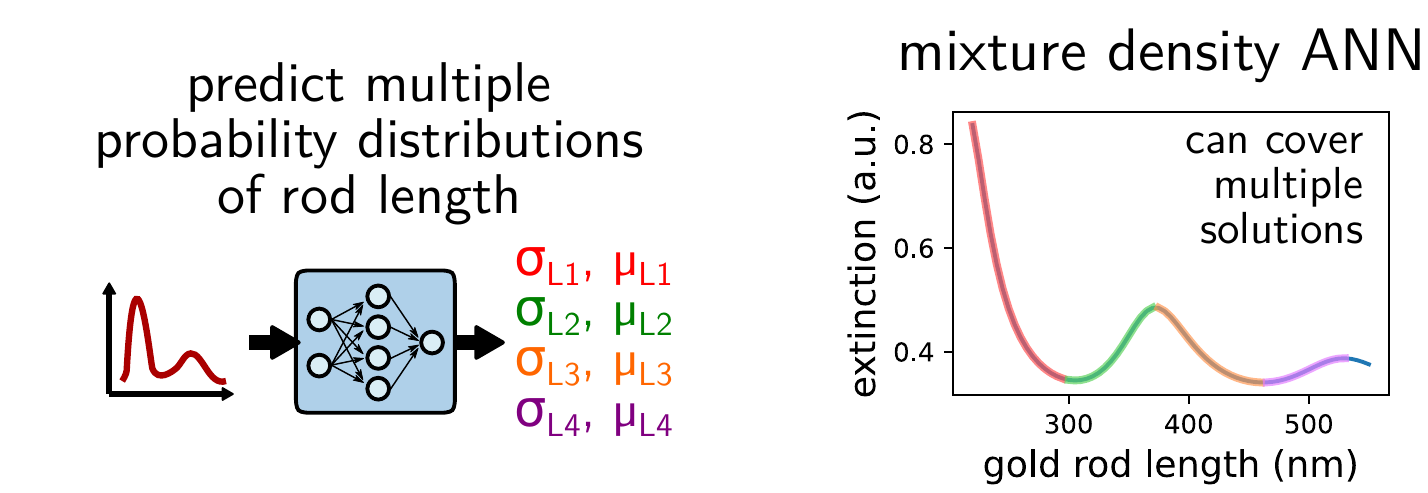}
		\end{center}
		
		Conditional generative adversarial networks (cGANs) or conditional [adversarial] autoencoders (c[A]AE) add a normal distributed latent vector (usually: ``$z$'') to the design target (``condition''), which encodes dynamically multiple possible solutions.
		Adversarial models furthermore use a trained loss, a so-called discriminator network (D), which tries to distinguish generated (fake) from training samples (true).
		\begin{center}
			\includegraphics[width=0.95\columnwidth]{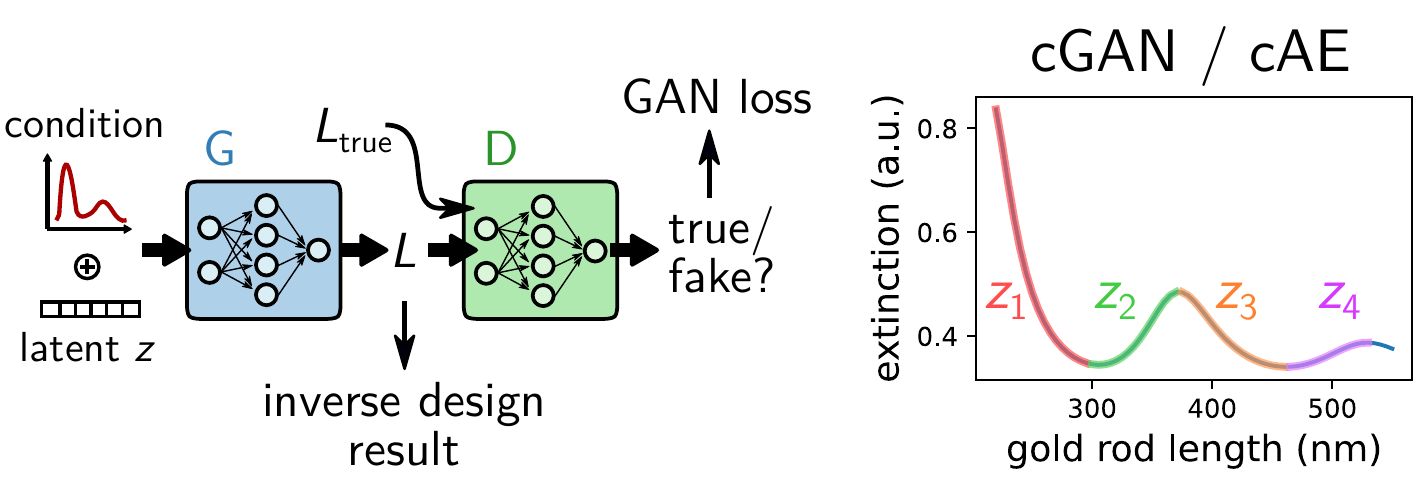}
		\end{center}

	\end{multicols}
\end{tcolorbox}

\subsection*{Deep learning forward solver}

ANNs have been successfully trained on the prediction of various physical quantities in nanophotonics.
Early works have proposed ANNs to create phenomenological models of non-linear optical effects or of optical ionization using experimental training data \cite{selleModelingLightmatterInteractions2007, selleModellingUltrafastCoherent2008}.
Recently, the idea has been picked up and it has been shown for instance that scattering and extinction spectra can be predicted with high accuracy \cite{malkielPlasmonicNanostructureDesign2018} and also that the phase can be included in the predictions \cite{anDeepLearningApproach2019}, which is important for nanostructures in metasurfaces.
The prediction of far-field observables can also be extended to include proximity effects in a dense metasurface, beyond the local phase approximation. 
The latter has been demonstrated by including the nearfield interactions with the nearest neighbor structures in the training data \cite{zhelyeznyakovDeepLearningAccelerate2020}.
The prediction of physical effects is not limited to extinction, transmission or other far-field effects. 
It has been shown that also nearfield effects can be approximated accurately, for instance around nanowires of complex shape \cite{liPredictingScatteringComplex2020}.

While networks that predict an observable like the scattering cross section perform usually very well within the range of their training data, such models often generalize rather poorly to cases outside the parameter range covered by the training data. 
The ANN acts then as universal function approximator, but it does not develop a deeper ``understanding'' of the underlying physics.
In order to alleviate this problem, it turned out to be helpful to provide the network with pre-processed data. 
For instance, instead of training an ANN with pure optical extinction spectra, So et al. \cite{soSimultaneousInverseDesign2019} trained their model using a decomposition in multiple electric and magnetic dipole resonances (ED, respectively MD), to predict the optical response of multi-material multi-shell nano-spheres. The approach is illustrated in figure \ref{fig:forward_models}a and has also been used to inverse design multi-shell spheres for Kerker-type directional scattering.
Using a  metallic grating as model example, Blanchard-Dionne and Martin demonstrated that a neural network that learns light-matter interaction through a representation as multiple Lorentz-oscillators generalizes about an order of magnitude better outside the training data range, compared to a predictor network based on the raw optical spectrum \cite{blanchard-dionneTeachingOpticsMachine2020} (see figure \ref{fig:forward_models}b).
Instead of predicting specific physical observables such as the extinction cross section, Wiecha et al. demonstrated that a network can learn a discrete dipole approximation of the electric polarization density inside a 3D nanostructure of arbitrary shape \cite{wiechaDeepLearningMeets2020}. 
The concept is depicted in figure \ref{fig:forward_models}c and allows to accurately derive manifold secondary quantities in the near- and far-field from a single generalized predictor neural network.

\subsection*{Forward predictor networks + evolutionary optimization}

In general, the greatest advantage of deep learning techniques as surrogate models for physics simulations is their tremendous evaluation speed. 
Once trained, an ANN delivers its prediction within fractions of milliseconds, which is usually orders of magnitude faster than a numerical simulation.
Therefore, replacing conventional physics simulations by surrogate ANNs is a natural solution to speed-up the inverse design of photonic nano-structures via global optimization heuristics \cite{zhuPhysicsconstrainedDeepLearning2019, chughSurrogateassistedEvolutionaryOptimization2020}.
This concept has recently been applied by several groups to the design of individual photonic nanostructures or metasurfaces \cite{campbellAdvancedMultiobjectiveSurrogateassisted2018, kaltMetamodelingHighcontrastindexGratings2019, gonzalez-alcaldeEngineeringColorsAlldielectric2020, pestourieActiveLearningDeep2020, hegdePhotonicsInverseDesign2020, kudyshevMachineLearningAssisted2020}.

However, while the approach can significantly accelerate heuristics-based inverse design, it remains an iterative approach requiring thousands of calls to the surrogate model as well as intermediate computation steps. 
Furthermore, the surrogate model represents only an approximation to the physical reality, introducing a systematic error. 
And even worse than that, it cannot be guaranteed that the surrogate model does not contain singular points of totally false solutions \cite{suOnePixelAttack2019}, to which the optimization algorithm may converge in the worst case scenario. 
Robust implementations therefore require a simulation-based fine-tuning procedure subsequent to the surrogate-based optimization run, which often relativizes the gain in speed \cite{jiangFreeFormDiffractiveMetagrating2019, trivediDatadrivenAccelerationPhotonic2019}. 
The same problem holds of course also for the here-after discussed ANN-only inverse design methods.

\subsection{Direct neural network inverse design}

As mentioned above, using forward ANNs as surrogate models for evolutionary optimization is computationally not the most efficient technique and bares the risk of converging to singular points of the surrogate model. 
In the recent past tremendous efforts have therefore been dedicated to the development of exclusively ANN-based inverse design schemes. 
The main obstacle which needs to be circumvented is the so-called ``one-to-many'' problem, which describes the fact that most inverse design problems are ambiguous, hence several non-unique solutions exist for the same design target.
In consequence a naive inversion of the ANN layout usually fails \cite{liuTrainingDeepNeural2018}, but several solutions have been developed to tackle the one-to-many problem. 
One possibility is the above described technique to use a forward-network as surrogate model, coupled to a global optimization algorithm. 
In this section we give a brief overview of pure neural network models to solve non-unique inverse problems. 
The different concepts are also schematized in Box~\ref{box:one_to_many}.

\refstepcounter{BoxCounter}\label{box:latent}
\begin{tcolorbox}[breakable, float=t, title={Box \arabic{BoxCounter}. Variational autoencoders and the latent space}]
	
	Variational autoencoders (VAEs) learn to compress information in a lower-dimensional \textit{latent space}, by being trained on a reconstruction task.
	
	\begin{center}
		\includegraphics[width=\columnwidth]{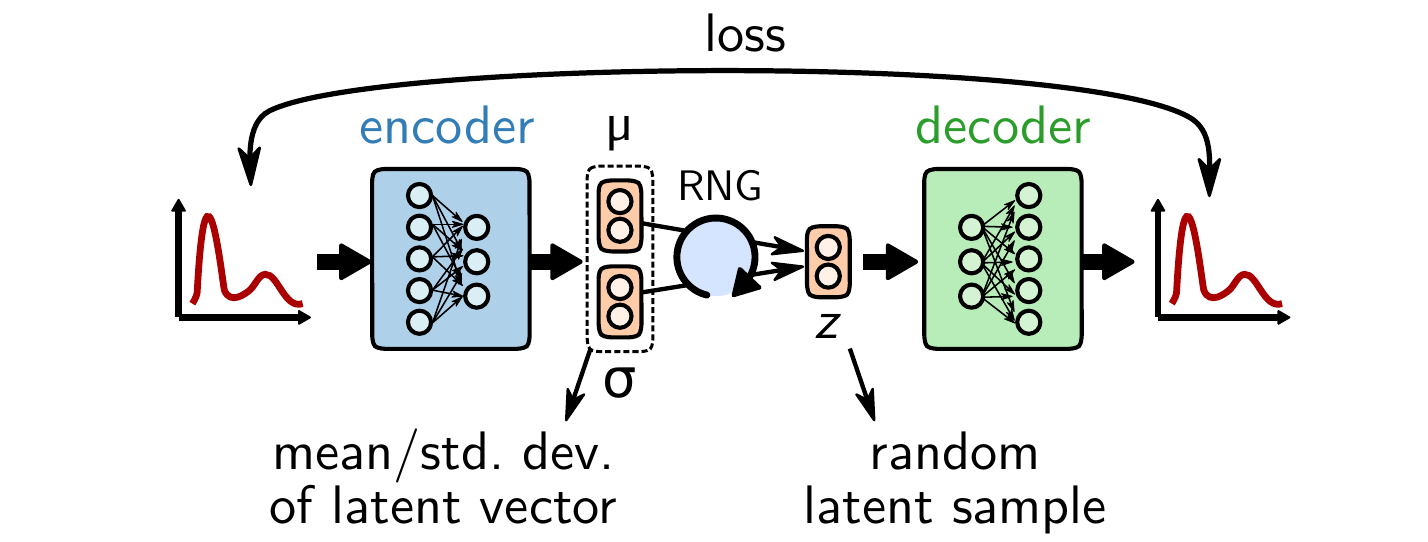}
	\end{center}
	
	In a VAE, forward propagation uses a random number generator (RNG) to draw samples $z$ with mean value $\mu$ and standard deviation $\sigma$. The random component ensures that the learned latent variables $z$ follow a normal distribution.
	However, gradient descent training requires analytical gradients, which cannot be back-propagated through the RNG. This is why a \textit{re-parametrization} into deterministic layers of $\mu$ and $\sigma$ is necessary \cite{kingmaIntroductionVariationalAutoencoders2019}.
	
	\begin{center}
		\includegraphics[width=0.85\columnwidth]{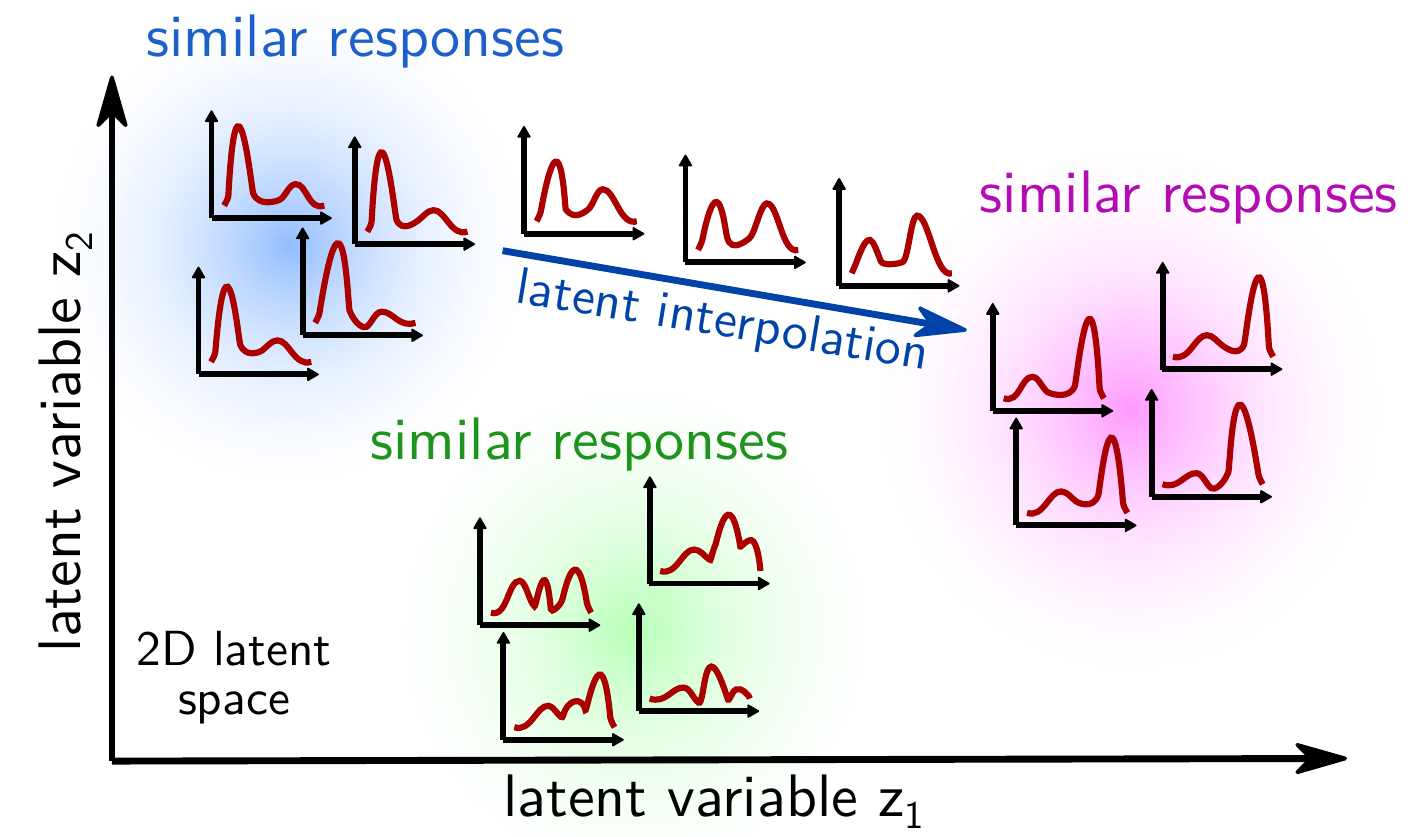}
	\end{center}
	
	By forcing the latent variables on a normal distribution, the trained VAE clusters similar inputs in the latent space. Furthermore transitions between solutions in latent space are smooth, which allows for example interpolation operations.
	
\end{tcolorbox}

A popular type of a stable inverse design network is the so-called \textit{tandem network} architecture \cite{malkielPlasmonicNanostructureDesign2018, liuTrainingDeepNeural2018, maDeepLearningEnabledOnDemandDesign2018, soSimultaneousInverseDesign2019, gaoBidirectionalDeepNeural2019}. 
In a tandem ANN a forward solver network is trained in a first step. 
The training of the actual inverse design network (the \textit{generator}) subsequently uses the fixed pre-trained forward model as a physics predictor to evaluate the inverse design output. 
In consequence, the loss function does not compare ambiguous design layouts but operates in the physics domain (comparing e.g. the extinction efficiency rather than the design parameters). 
In this way, different design parameters which lead to a similar physical response no longer confuse the ANN, all correct solutions to a given design problem yield a positive training feedback.

Another model that circumvents the one-to-many problem is the conditional Generative Adversarial Network (cGAN) \cite{liuGenerativeModelInverse2018, soDesigningNanophotonicStructures2019, jiangFreeFormDiffractiveMetagrating2019, anMultifunctionalMetasurfaceDesign2020, mallFastDesignPlasmonic2020}. 
A cGAN takes as input not only the design target but also an additional ``latent vector'', which is a normally distributed sequence of random values. 
The network then learns to use different values of the latent vector to address the distinct non-unique solutions. 
In addition to the introduction of a latent vector, a further peculiarity of cGANs is their loss function, which is is a discriminator network that tries to distinguish generated solutions from real ones, and which is also subject to training.
During training, the cGAN loss function hence evolves together with the ANN, which allows ideally a better convergence. 
It is worth to note, that it is a delicate task to tune the network and training hyperparameters in GANs such that the learning converges. The training of both, generator network and discriminator network needs to evolve in a balanced way for the adversarial loss function to work efficiently.

A further type of one-to-many solving networks are conditional adversarial or conditional variational autoencoders (cAAEs, cVAEs) \cite{liuHybridStrategyDiscovery2020, maProbabilisticRepresentationInverse2019a, shiMetasurfaceInverseDesign2020, kudyshevMachineLearningAssisted2020, maDataefficientSelfsupervisedDeep2020}. 
Those are usually symmetric models that take the physical response as input, which they try to identically reconstruct at their output layer.
In a conditional autoencoder, a bottleneck layer is placed in the ANN center. This bottleneck contains the design parameters on the one hand (as it is the case in a tandem-network), but on the other hand an additional latent vector is appended to the design parameters. 
Like in the cGAN, the latent vector can be used by the ANN to address potential multiple solutions. Unlike in the tandem network the forward model is trained simultaneously with the generator.
Conditional autoencoders can be seen as a mixture of a tandem network and a cGAN. 
For a short explanation of the basic idea behind VAEs and the meaning of the latent space, see also Box~\ref{box:latent}.

For completeness we want to mention also work on reinforcement learning for iterative design optimization, where the neural network learns to behave as an iterative optimization algorithm. 
The expectation is that the ANN can adapt its optimization strategy specifically to the given problem and hence outperform conventional heuristic algorithms \cite{badloeBiomimeticUltrabroadbandPerfect2020, wangAutomatedMultilayerOptical2020}.

The above discussed models have been quite successfully used for manifold inverse design problems in nano-photonics. 
Figure~\ref{fig:ML_inverse_design}a shows an example of multi-mode interference devices (MMIs) designed by a tandem-ANN. 
MMIs are large waveguides that support many modes and that can have several inputs and outputs (here $3\times 3$). 
The here shown MMIs are patterned with small perturbations in order to obtain specific light-routing properties. 
The tandem ANN has been trained to design perturbation patterns which produce arbitrary transmission states. 
This allows for instance to define MMI patterns which swap a pair of the $3\times 3$ in- and output paths, while one of the transmission channels remains constant, as demonstrated in figure~\ref{fig:ML_inverse_design}a (i) and (ii) \cite{dinsdaleDeepLearningEnabled2021}.
Figure~\ref{fig:ML_inverse_design}b shows a meta-surface which acts as a flat lens with two focal spots, designed by a variant of a cGAN network \cite{anMultifunctionalMetasurfaceDesign2020}.
Other examples are the design of chiral plasmonic structures \cite{maDeepLearningEnabledOnDemandDesign2018, ashalleyMultitaskDeeplearningbasedDesign2020}, 
dielectric structures \cite{trisnoApplyingMachineLearning2020},
multi-shell nano-spheres \cite{soSimultaneousInverseDesign2019, peurifoyNanophotonicParticleSimulation2018}, invisibility cloaks \cite{sheverdinPhotonicInverseDesign2020, blanchard-dionneSuccessiveTrainingGenerative2021, chenPhysicsinformedNeuralNetworks2020}, or metasurface design \cite{sajedianDoubledeepQlearningIncrease2019, phanDeepLearningInverse2020, yeungDesigningMultiplexedSupercell2020}.

\subsection{Strategies to improve neural-network inverse design}

\begin{figure*}[t!]
	\centering{
		\includegraphics[width=0.9\linewidth]{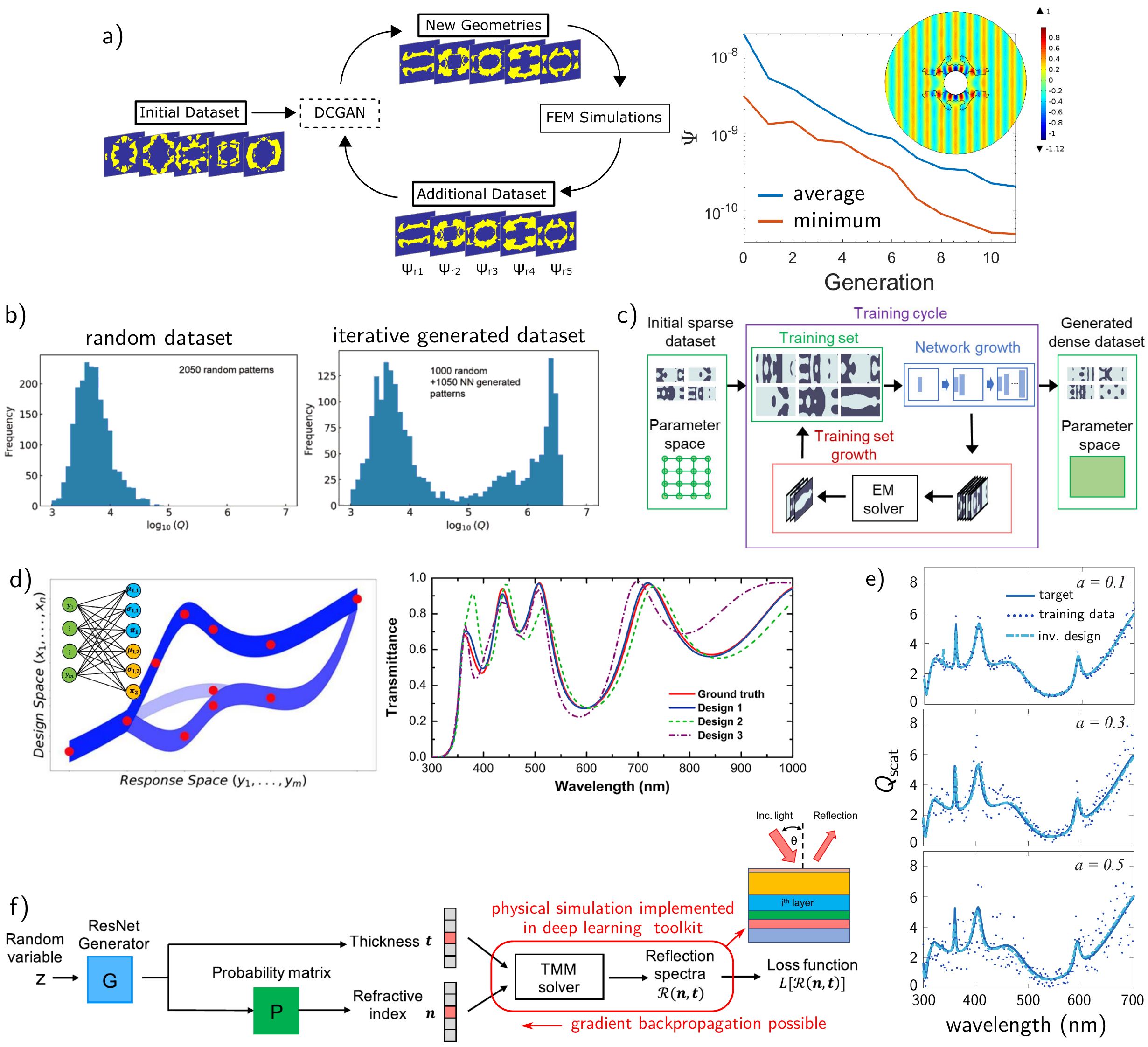}}
	\caption{
		Concepts to improve common shortcomings of inverse design ANNs.
		(a) iterative training data generation, in which a network learns from its own errors, here applied to the inverse design of an invisibility cloak device. Adapted from \cite{blanchard-dionneSuccessiveTrainingGenerative2021}, copyright (2021) Optical Society of America.
		(b) Comparison of the Q-factors for photonic crystal cavities in a random dataset (left) and in an iteratively generated dataset after the first iteration (right). Adapted from  \cite{asanoIterativeOptimizationPhotonic2019}, copyright (2019) de Gruyter.
		(c) together with the training data, the network complexity can be progressively growing, allowing even better performance by successive learning of smaller features. Reprinted with permission from \cite{wenRobustFreeformMetasurface2020}. Copyright (2020) American Chemical Society.
		(d) mixture density ANN which represents multiple solutions with Gaussian probability distributions to find several non-unique solutions to ambiguous problems. The shown example deals with the spectral design of a multi-layer stack.  Adapted with permission from \cite{unniDeepConvolutionalMixture2020}. Copyright (2020) American Chemical Society. 
		(e) de-noising inverse ANN as robust approach for training on noisy data (noise parameter $a$ increasing from top to bottom). Adapted from \cite{huRobustInversedesignScattering2019}, copyright (2019) Optical Society of America.
		(f) ``GLOnet'': Inverse design ANN using a transfer-matrix model loss for reflectivity and transmission spectra optimization of multi-layer stacks. 
		Adapted from \cite{jiangMultiobjectiveCategoricalGlobal2020}, copyright (2020) de Gruyter.
	}
	\label{fig:improve_inv_design}
\end{figure*}

Data-driven inverse design has the important drawback that the accuracy of the model is first of all limited by the quality of the data and an interpolation error between the data-samples is introduced by the ANN. 
Early works on inverse design therefore reported rather qualitative agreement, but relatively large quantitative inaccuracies.
Therefore, in the recent past remarkable efforts have been put in developing methods to improve neural network inverse design. 
In this section we want to provide an overview over the most successful concepts.
In general, two main constituents offer the largest potential for optimization: The training data and the neural network model.

\subsection*{Improving the data quality}

As mentioned before, many ANN models do actually generalize relatively poorly to cases outside the parameter range of the training data. 
They act mainly as generalized function approximators, hence they interpolate very efficiently to fill the gaps in the training data, while their extrapolation capability remains limited.
But also the interpolation risks to be unsatisfactory if the physical model underlying the training data has sharp features such as high quality factor resonances. 
If the training data does not contain a sufficient number of such resonant cases, there is a high risk that those features will be very poorly approximated by an ANN. 

To tackle this problem, training data can be generated using an optimization algorithm to produce specific responses for the dataset \cite{sheverdinPhotonicInverseDesign2020, dinsdaleDeepLearningEnabled2021}. 
In case of many free parameters this procedure is time-consuming, therefore recently the idea of iterative training data generation has emerged \cite{dinsdaleDeepLearningEnabled2021, sheverdinPhotonicInverseDesign2020, blanchard-dionneSuccessiveTrainingGenerative2021, asanoIterativeOptimizationPhotonic2019, pestourieActiveLearningDeep2020, kudyshevMachineLearningAssisted2020}. 
The principal idea is depicted in figure~\ref{fig:improve_inv_design}a. 
An initial dataset is generated traditionally via a randomized procedure, on which an inverse design ANN is trained.
This network is subsequently used to construct devices based on realistic design targets, but these designs are likely to be rather mediocre as the initial ANN performs relatively poor.
Now, the true physical response of these mediocre ANN designs is calculated in another run of numerical simulations, and these samples are appended to the training data. 
The generator ANN is then trained again on the now extended training data and the generative cycle is repeated.
In this way, the neural network can literally learn from its previous mistakes and its performance on the specific design task will significantly improve.
Fig.~\ref{fig:improve_inv_design}a shows the example of an optical cloak design problem, for which the inverse design accuracy could be improved by more than one order of magnitude thanks to iterative training \cite{blanchard-dionneSuccessiveTrainingGenerative2021}.
To visualize the evolution of the training data quality, figure~\ref{fig:improve_inv_design}b shows the statistical distributions of resonator quality factors in a fully random dataset of photonic crystal cavities (left) compared to a dataset after one iteration of iterative training (right) \cite{asanoIterativeOptimizationPhotonic2019}. 
The lack of resonant geometries in the randomly generated dataset is evident. 
Despite those solutions not being present in the initial dataset, the ANN managed to conjecture a certain amount of resonant cases, improving the training data for the second iteration. 
By repeating the procedure the training data increasingly contains resonant geometries, which consequently allows the ANN to inverse design close-to-optimum solutions.
Another positive side-effect specifically in tandem networks is that iterative training simultaneously improves the accuracy of the forward network \cite{dinsdaleDeepLearningEnabled2021}.
A recent work showed that an even better design performance can be achieved by iteratively increasing the network complexity together with a successive augmentation of the training-data, as depicted in figure~\ref{fig:improve_inv_design}c \cite{wenRobustFreeformMetasurface2020}.

\begin{figure*}[t!]
	\centering{
		\includegraphics[width=0.9\linewidth]{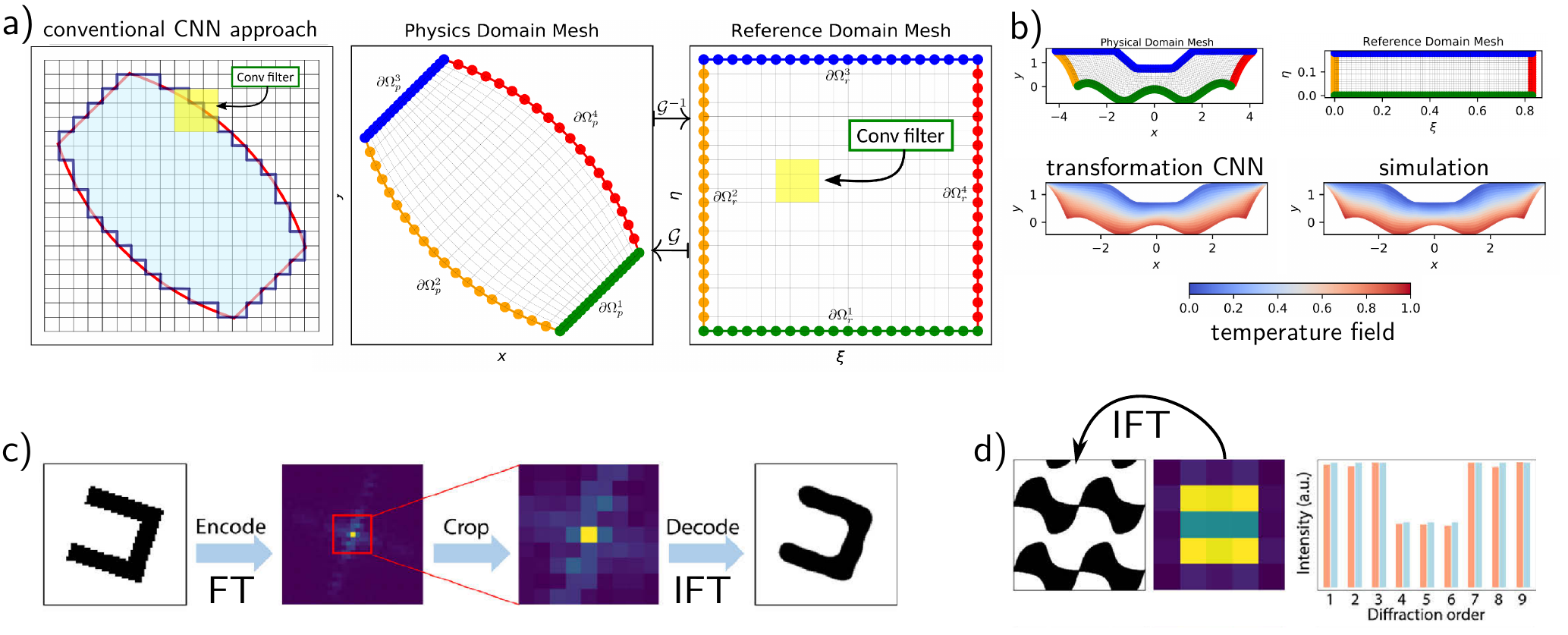}}
	\caption{
		Examples of input data pre-processing for optimized physics domain representation. 
		(a-b) deep learning on irregular grids via coordinate transform (a) which is implemented within the deep learning toolkit to allow fast gradient calculations through the coordinate system transformation. (b) the transformation allows to efficiently train networks on complex shaped physical domains. Adapted with permission from \cite{gaoPhyGeoNetPhysicsinformedGeometryadaptive2020}, copyright (2020) Elsevier.
		(c) Data encoding and compression using a topology description based on low-frequency Fourier components, which allows data-efficient treatment of complex shapes, here for example a free-form metagrating. Adapted from \cite{liuTopologicalEncodingMethod2020}, copyright (2020) Optical Society of America.
	}
	\label{fig:improved_data_structures}
\end{figure*}

\refstepcounter{BoxCounter}\label{box:wisdom_of_many}
\begin{tcolorbox}[breakable, float=t, title={Box \arabic{BoxCounter}. ``Wisdom of the many''}]
	
	\textit{Wisdom of the many} or also \textit{wisdom of the crowd} denotes the procedure of training multiple neural networks on the same data, each ANN with random initialization. 
	We illustrate the idea by the example of an optical spectrum predictor network:
	
	\begin{center}
		\includegraphics[width=0.99\columnwidth]{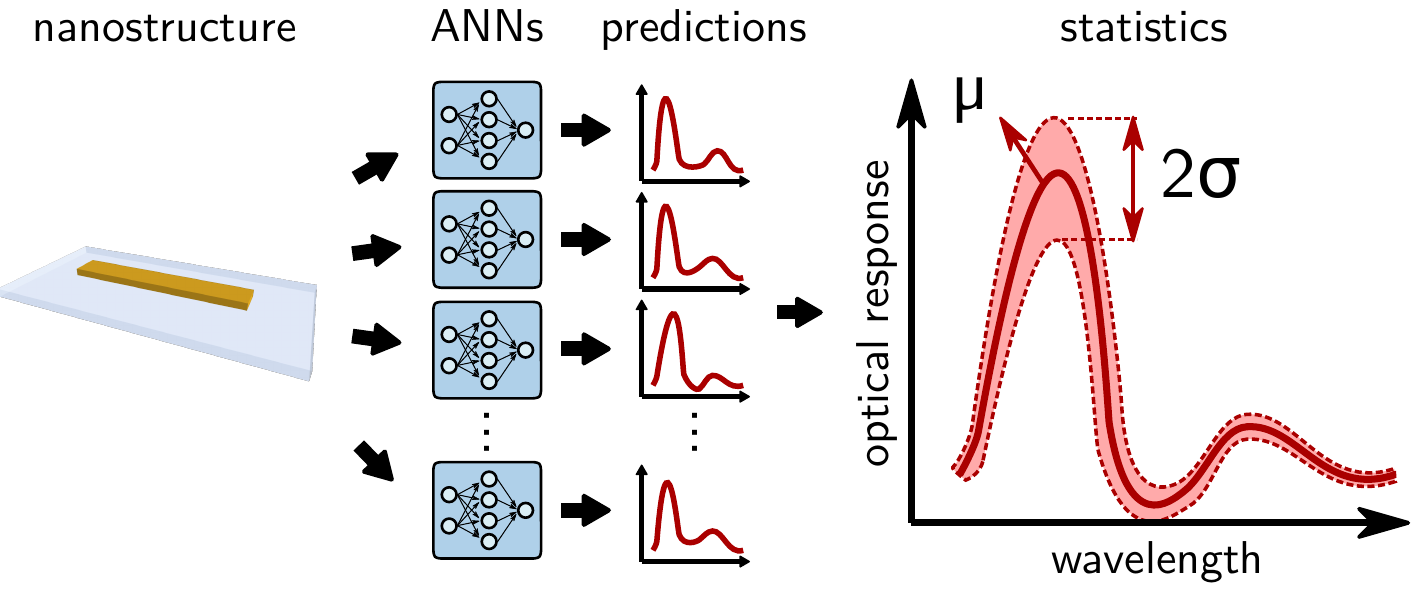}
	\end{center}
	
	While this approach adds a significant computational cost (training several networks), the mean \textmu\ of $N$ independent predictions provides a $\sqrt{N}$ times smaller statistical error compared to using a single ANN.
	Furthermore, the standard deviation $\sigma$ of multiple predictions can be used to assess the credence of the ANN output.
	
\end{tcolorbox}

An obvious drawback of iterative procedures is their computational cost. Data generation is usually slow and the expensive network training needs to be repeated several times on increasing amounts of training samples. 
Several suggestions have been made to accelerate the convergence of iterative data generation in order to reduce the number of cycles.
For instance by training several networks, the statistics from multiple predictions can be used to assess the quality and the uncertainty of the ANN output (``wisdom of the many'' \cite{wangMassiveComputationalAcceleration2019}, see also Box~\ref{box:wisdom_of_many}). This information can be exploited to choose only the best new solutions for re-simulation and insertion into the expanded training data, which reduces the number of expensive physics simulations \cite{pestourieActiveLearningDeep2020}.
Similarly, an evolutionary optimization algorithm might be coupled to a generative ANN in the iterative cycle to further specialize the training data with regards to the anticipated optimization target \cite{kudyshevMachineLearningAssisted2020}.
A drawback of such training-data optimization strategies is a risk of over-specializing the network to optimum cases and losing its capability to generalize to arbitrary situations. 
Therefore care needs to be taken that the training data remains sufficiently diverse.

\subsection*{Physics model based loss function}

A similar, yet somehow more radical concept is to not use a fixed set of training data at all but instead implement a loss function based on a physical model within the framework of the machine learning toolkit.
Such an approach has been illustrated recently by the example of inverse designing multi-layer thin-film stacks for specific reflection and transmission spectra  \cite{jiangMultiobjectiveCategoricalGlobal2020}.
As highlighted by a red box on the right in figure~\ref{fig:improve_inv_design}f, a transfer-matrix method (TMM) has been implemented directly in the deep learning toolkit as a loss function. 
In consequence, error backpropagation is possible through the TMM solver, and the network can be trained without an explicit dataset.
The loss function in this so-called ``GLOnet'' is used to optimize the transmission and reflection spectra of a multi-layer stack with respect to a design target.
It is worth mentioning that the GLOnet learns to optimize a single design target, hence in principle the training of the network takes the place of a conventional global optimization algorithm run (hence its name ``GLOnet''). 
The authors of Ref.~\cite{jiangMultiobjectiveCategoricalGlobal2020} claim that the training dynamics allow their GLOnet to ideally adapt its optimization scheme to each problem, resulting in better and faster convergence compared to hard-coded optimizers.
The same authors have generalized their concept to a somehow more flexible inverse network called ``conditional GLOnet'', using an iterative training scheme instead of a fully differentiable physics loss function. 
For the training, gradients of the design efficiency are calculated via adjoint simulations and re-injected for backpropagation through the network \cite{jiangGlobalOptimizationDielectric2019}. 
The conditional GLOnet is conceptually similar to a Pareto optimization in which a set of optimum solutions for a multi-objective problem is calculated \cite{debMultiobjectiveOptimizationUsing2001}.
While the specific solving of a single problem is intentional in Refs.~\cite{jiangMultiobjectiveCategoricalGlobal2020, jiangGlobalOptimizationDielectric2019}, as already mentioned before over-specialization is an inherent danger of all iterative data-generation methods.

Another concept to replace the dataset by a direct evaluation of a physics model has been demonstrated for the Helmholtz equation, by developing a loss function which directly evaluates this partial differential equation (PDE). 
Such ANN model is called a ``physics informed neural network'' (PINN). 
In case of a Helmholtz-PINN, the network learns to directly solve the wave equation in frequency domain. 
The inverse design target is then implemented as a boundary condition matching problem \cite{fangDeepPhysicalInformed2020, chenPhysicsinformedNeuralNetworks2020}.
As in the GLOnet case, also such PINN inverse design requires a new training run for each optimization target.
PINNs will be discussed in more detail later in this review.

\subsection*{Sophisticated ANN models} 

The second main lever allowing for performance optimization of inverse design ANNs is the neural network model itself.
It has been proven helpful to adopt recent findings in the research on optimum network layout for deep learning.
For instance, if applicable the ``U-Net'' architecture \cite{ronnebergerUNetConvolutionalNetworks2015} offers much better training convergence and generalization capacity than standard convolutional neural networks -- even in cases where its particularly efficient segmentation capabilities are not required \cite{borhaniLearningSeeMultimode2018, wiechaDeepLearningMeets2020}.
Furthermore so-called residual blocks, or ResNets \cite{szegedyInceptionv4InceptionResNetImpact2016}, should be adopted whenever possible. 
Residual blocks are characterised by their skip connections which avoid the vanishing gradient problem, allowing the training of very deep network layouts.

\begin{figure*}[t!]
	\centering{
		\includegraphics[width=0.9\linewidth]{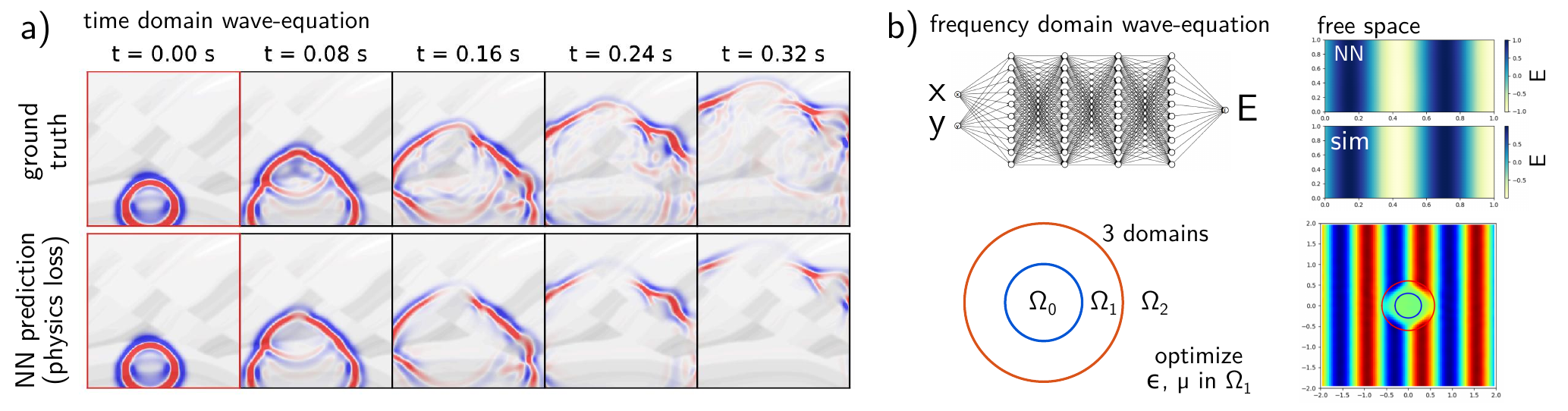}}
	\caption{
		Physics informed neural networks (PINNs) for nano-optics.
		(a) PINN for solving the wave equation in time domain. Adapted with permission from \cite{moseleySolvingWaveEquation2020}.
		(b) top: solving the Helmholtz equation (frequency domain), bottom: using the PINN for inverse design of the permittivity distribution in domain $\Omega_1$ for an invisibility cloak application. Adapted with permission from \cite{fangDeepPhysicalInformed2020}, copyright (2020) IEEE.
	}
	\label{fig:PINNs}
\end{figure*}

In addition to the application of general ``best-practice'' ANN design rules, problem-specific tailoring of the network layout can be very favorable for optimum inverse design performance.
For instance, to tackle the one-to-many problem, ``multi-branch'' or ``mixture density'' ANNs can be applied in addition to the above named network architectures. 
The concept is based on representing the design parameters in a ``modal'' representation as multiple Gaussian distributions, where each of the Gaussian distributions describes a possible solution to an ambiguous problem (see also box~\ref{box:one_to_many}).
This concept has been proposed some time ago for microwave device inverse design \cite{kabirNeuralNetworkInverse2008, zhangMultivaluedNeuralNetwork2018} and was recently adapted to nano-photonics \cite{unniDeepConvolutionalMixture2020, luoProbabilityDensityBasedDeepLearning2020} (see also figure~\ref{fig:improve_inv_design}d). 
The advantage is that the network can in principle deliver all possible solutions together with a weight for their respective priorities. 
A drawback of the approach is that the approximate number of non-unique solutions needs to be known in advance.

Another recent proposition to optimize inverse networks specifically for noisy situations like in experiments is the implementation of concepts from machine learning based image denoising \cite{xieImageDenoisingInpainting2012}. 
As shown in figure~\ref{fig:improve_inv_design}e, Hu et al. added artificial noise on training data and could demonstrate that a denoising network based inverse ANN offers a very robust performance even when trained on very noisy data \cite{huRobustInversedesignScattering2019}. 
This opens promising perspectives for experimental applications.

\subsection*{Reformatting the input data}

Apart from optimizing the network model and generating training data of high quality, the \textit{format} of the inputs and outputs of a neural network can play a decisive role in whether the ANN manages to ``understand'' the data or not.
An example is illustrated in figure~\ref{fig:improved_data_structures}a, where a physical problem is to be solved on a non-Cartesian coordinate domain. 
On 2D problems such as the here shown one, typically convolutional neural networks (CNNs) are most efficient. 
However, as can be seen in the leftmost panel, the imposed discretization on a square mesh is very poor. This holds in particular for the domain borders. 
Gao et al. \cite{gaoPhyGeoNetPhysicsinformedGeometryadaptive2020} proposed therefore to apply a transformation of the coordinate system from the physics domain to the CNN reference domain prior training. 
As illustrated in figure~\ref{fig:improved_data_structures}b by the example of solving the heat equation, this additional pre-processing allows to successfully apply ANNs to very complex non-uniform physical domains.

The problem of discretization can also be alleviated by applying a topology encoding procedure, for instance via Fourier transformation \cite{liuTopologicalEncodingMethod2020}. 
The idea is illustrated in figure~\ref{fig:improved_data_structures}c-d.
Such encoding can allow not only to describe geometries with odd shapes without restrictions due to discretization, it allows furthermore to condense the information to a low-dimensional space, which is helpful to reduce ANN complexity and furthermore advantageous in preventing overfitting.


\subsection*{Other concepts} 

Further possibilities to improve the quality of ANN based inverse design are to use the ANN only as a first step for a rough estimate, and apply a conventional iterative approach in a subsequent refinement step. 
Heuristic optimization algorithms usually benefit strongly from a good initial guess \cite{jiangFreeFormDiffractiveMetagrating2019}.
Another recent proposition is to use a forward neural network purely as an ultra-fast physics predictor to construct a huge look-up table \cite{nadellDeepLearningAccelerated2019}.
Using a well trained forward network, a look-up table can be created which covers the entire parameter space at a very fine resolution, impossible to achieve with conventional numerical methods. Appropriate solutions to specific problems can subsequently be searched in this database.
Transfer learning has also been recently applied to nano-optics problems to improve ANN performance if only small amounts of data exist \cite{quMigratingKnowledgePhysical2019}. 
For instance experimental data is often expensive, but the situation can be improved by training an ANN first on simulated data, and subsequently specialize the pre-trained network via transfer learning on the experimental dataset \cite{narhiMachineLearningAnalysis2018}.

\subsection{Heuristics vs. deep learning -- a critical comparison}

It is of utmost importance to emphasize that a data-driven inverse design technique can never outperform an iterative method if it is based on the same simulation model used for training data generation. At least not if no time constraint is set for the iterative optimization. 
Well trained and optimized data-driven ANNs usually produce errors in the order of a few percent \cite{liPredictingScatteringComplex2020, wiechaDeepLearningMeets2020}. 
Furthermore, it is virtually impossible to completely suppress outliers in the network predictions \cite{suOnePixelAttack2019}. At the singular points the error of the ANN can be orders of magnitude higher. 
It is thus a delicate task to assess whether a prediction is valid or rather the result of a singularity in the ANN.

While recently some sophisticated training techniques were presented that are capable to train ANNs for performances similar to conventional inverse optimization, they are either still considerably constrained or the high accuracy has a severe impact on the computational cost. Examples are physics-loss based inverse ANNs or networks based on progressive-complexity training schemes \cite{wenRobustFreeformMetasurface2020, jiangMultiobjectiveCategoricalGlobal2020}.
The model described in Ref.~\cite{jiangMultiobjectiveCategoricalGlobal2020} for example is constrained to a simple transfer-matrix description of a multi-layer system as well as to the inverse design of a single optimization target.

The fact that ANNs always introduce an additional error is inherent to the data-driven nature of machine learning, which implies that a ML model can never outperform the accuracy of the simulations used to create the dataset or the model defining the training loss. 
On the other hand, once trained ANN techniques can offer extreme speed-up of the inverse design, generally many orders of magnitude faster than iterative approaches based on numerical simulations. 
It is not unusual that milliseconds stand against hours or even days. 
This is a marvellous advantage and often well worth to accept the reduced accuracy of ANN based techniques.
In daily applications a few percent error might actually not even matter too much, in particular when compared to the typical magnitude of inaccuracies in fabrication. 

On the other hand, concerning the inverse design speed it is important to remember that the ultra-fast predictions require a fully trained neural network. This implies the computationally highly demanding data generation as well as the very expensive training of the ANN.
In many situations, conventional global optimization is in sum actually computationally cheaper.
In conclusion, deep learning based inverse design is mainly interesting for applications which require a large number of repetitions of similar design tasks, or that rely on ultimate speed for the design generation.

\section{Beyond inverse design}

The second part of this review is dedicated to applications of deep learning in nano-photonics ``beyond inverse design''. 
We give an overview on physics informed neural networks, we present recent work on ANNs for physics interpretation and knowledge discovery as well as experimental applications.

\subsection{Physics informed neural networks: Solving PDEs}

Most machine learning applications in physics aim to predict derived observables such as transmittance or extinction cross sections. 
In contrast, the idea of \textit{physics informed neural networks} (PINNs) is to train an ANN to directly predict the solution of a [partial] differential equation ([P]DE).
While this would be also possible using a dataset of pre-calculated solutions, the particularity of PINNs is that instead of using a loss function for data-comparison like MSE, the PINN-loss implements an explicit evaluation of the PDE. 
In consequence, no pre-calculated training data is required for training.
For the PINN-loss, the PDE derivatives of the ANN-predicted observables are directly implemented in the respective deep learning toolkit. 
Thus, the PINN-loss can be seen as a consistency check for the predicted solution.
Because modern deep learning toolkits offer powerful automatic differentiation functionalities, error backpropagation through the PINN-loss remains possible and the ANN can be efficiently trained without data.

\begin{figure*}[t!]
	\centering{
		\includegraphics[width=0.95\linewidth]{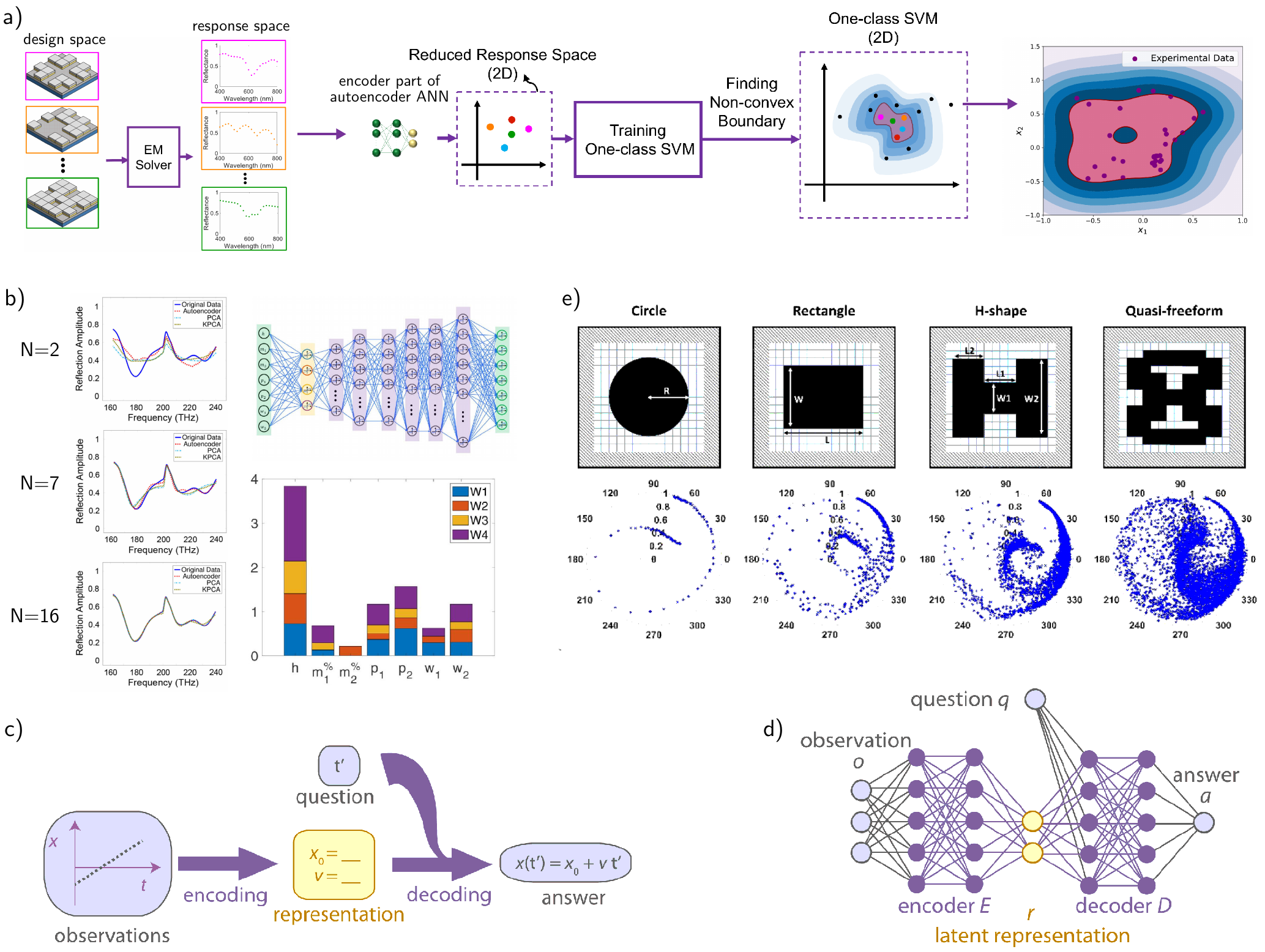}}
	\caption{
		Examples of ``knowledge discovery'' through machine learning.
		a) The feasibility of a physical response by a defined geometric model can be assessed by a dimensionality reduction through an autoencoder neural network and subsequent non-convex hull determination. Adapted from \cite{kiarashinejadKnowledgeDiscoveryNanophotonics2020}, copyright (2019) the authors.
		b) Study of the impact of the number of bottleneck neurons $N$ (left spectra) as well as of nanostructure design variations on the activation of the bottleneck neurons (W1-W4 in case $N=4$, yellow neurons in the top right panel). 
			This analysis allows to assess the physical importance of individual design parameters and reveals information about the complexity of the optical response. Adapted with permission from \cite{kiarashinejadDeepLearningReveals2019}, copyright (2019) John Wiley and Sons.
		(c-d) By mimicking the human approach of interpreting and modelling physical observations (c), a conditional encoder-decoder network (d) can be used to discover implicit physics concepts from data. Reprinted with permission from \cite{itenDiscoveringPhysicalConcepts2020}, copyright (2020) APS.
		e) Exploiting the high speed of a physics predictor network permits a systematic analysis of the achievable phase and intensity variations in metasurface constituent design. Adapted from \cite{anDeepLearningModeling2020}, copyright (2020) Optical Society of America.
	}
	\label{fig:knowledge_discovery}
\end{figure*}

This concept has been first proposed in 2019 by Raissi et al. \cite{raissiPhysicsinformedNeuralNetworks2019} and has since then attracted a great deal of attention across countless research communities in physics such as fluid mechanics \cite{raissiPhysicsinformedNeuralNetworks2019, raissiHiddenFluidMechanics2020}, thermodynamics \cite{gaoPhyGeoNetPhysicsInformedGeometryAdaptive2020} or geophysics \cite{moseleySolvingWaveEquation2020}.
Compared to data-based ANNs, the accuracy of PINNs is in general significantly higher.
On the other hand, because PINNs evaluate the underlying PDE ``point-by-point'', they are usually slower than conventional data-based models. Since the latter work on physical observables, it is easier to predict higher-dimensional data-structures at a time, making better use of the massively parallel computing architectures of modern GPUs. 
Nevertheless, PINNs are usually orders of magnitude faster than numerical PDE solvers.

Applications to nano-photonics are still scarce. Recently Moseley et al. demonstrated that PINNs are capable to accurately solve the wave equation in time domain \cite{moseleySolvingWaveEquation2020}.
An example is shown in figure~\ref{fig:PINNs}a, demonstrating seismic wave propagation through an inhomogeneous medium at successive snapshots in time. As can be seen, the PINN is capable to predict the evolution of the wave propagation even in a complex environment.
While Ref.~\cite{moseleySolvingWaveEquation2020} treats shock waves in geophysics, the problem is conceptually identical to the wave equation in electrodynamics.

Depicted in figure~\ref{fig:PINNs}b, Fang and Zhan recently demonstrated that a PINN can accurately solve the Helmholtz equation, describing wave propagation in frequency domain \cite{fangDeepPhysicalInformed2020}. 
They found that sinusoidal activation functions are the most adequate choice to solve a differential equation with time-harmonic solutions. 
By formulating the inverse design as a boundary condition matching problem, it was possible to use the Helmholtz-PINN for the design of an optical cloak, as illustrated in the bottom of figure~\ref{fig:PINNs}b.
A similar frequency-domain PINN has been proposed for the homogenization of optical metamaterials \cite{chenPhysicsinformedNeuralNetworks2020}.
A disadvantage of PINNs is that the environment needs to be defined at the training stage and hence a new network needs to be trained if the boundary conditions change. 
Each PINN-based inverse design therefore involves a new training procedure, comparable with conventional iterative techniques, which is evidently much slower than ``direct'' inverse-ANN models.
Conceptually related to PINNs is also the so-called ``GLOnet'', which is discussed in more detail above \cite{jiangMultiobjectiveCategoricalGlobal2020} (see also figure~\ref{fig:improve_inv_design}f).

\subsection{Interpretation of physical properties}

\begin{figure*}[t!]
	\centering{
		\includegraphics[width=0.9\linewidth]{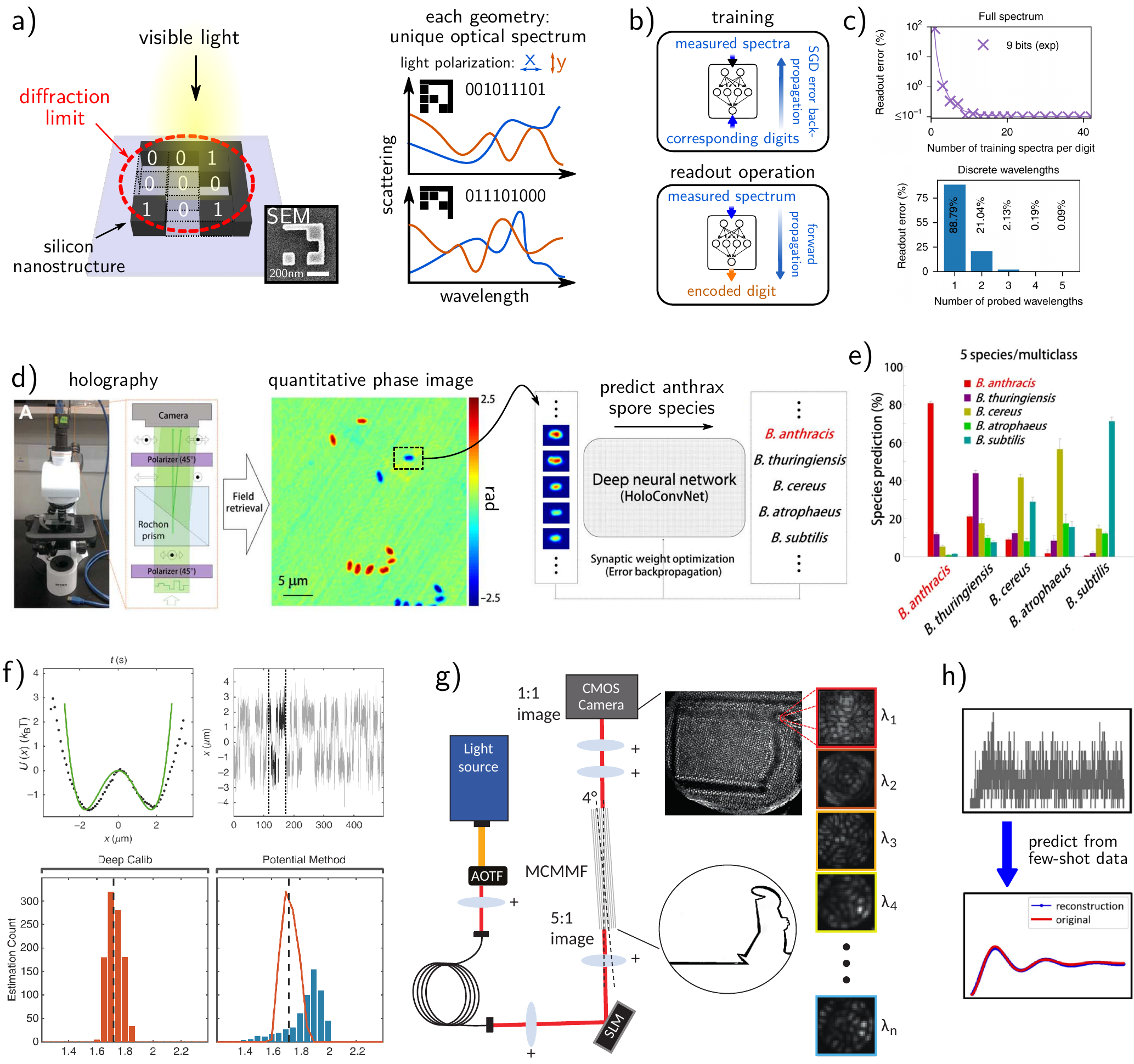}}
	\caption{
		Examples of ML applications in experimental data interpretation.
		(a-c) ANN used to decode information from optical information storage via a spectral scattering analysis from sub-diffraction small nanostructures. (a) Each bit sequence is encoded by a specific geometry which is designed such that it possesses a unique scattering spectrum. (b) A neural network is trained on a large amount of spectra such that it learns to decode noisy spectra of formerly not seen structures. (c) Even if only few wavelengths are probed, the readout accuracy of the network is excellent. Adapted with permission from \cite{wiechaPushingLimitsOptical2019}, copyright (2019) Springer Nature.
		(d-e) Holographic anthrax spore classification via holography microscopy. A machine learning algorithm is trained on phase images of different spore species, as depicted in (d). The neural network is capable to classify 5 different Anthrax species with a very high accuracy. Adapted from \cite{joHolographicDeepLearning2017}, copyright (2017) the authors.
		(f) Microscopy force field calibration (top left, green line: trapping potential, dots: reconstructed potential). 
		Evaluation of $U(x)$ via ANN-based analysis of Brownian motion from undersampled statistical data (top right). Comparison of reconstruction fidelity of ANN (bottom left) and conventional method (bottom right). Ground truth is indicated by a black dashed line. Adapted from \cite{argunEnhancedForcefieldCalibration2020}, copyright (2020) the authors.
		(g) ANN enabled real-time hyper-spectral image reconstruction from speckle patterns produced by a multicore multimode fiber bundle (MCMMF). The technique exploits the wavelength-dependence of the speckle patterns. Adapted from \cite{kurumDeepLearningEnabled2019}, copyright (2019) Optical Society of America.
		(h) Scheme depicting the use of machine learning for statistics reconstruction of few-shot data acquisitions. Reprinted from \cite{cortesAcceleratingQuantumOptics2020}, with the permission of AIP Publishing.
	}
	\label{fig:experimental_applications}
\end{figure*}

In this section we will review recent approaches to extract information and correlations from deep learning models in order to reveal physical insights.

There is on the one hand the possibility to use deep learning models for dimensionality reduction. 
Figure~\ref{fig:knowledge_discovery}a shows a work of Kiarashinejad et al. in which the number of discrete values in reflectance spectra from a set of electrodynamical simulations is reduced from 200 to 2 via an unsupervised autoencoder ANN.
In a second step, the non-convex hull of the compressed responses is calculated, which represents the region in the 2D compressed space containing all encoded points. 
This region allows to assess the range of accessible physical responses within the allowed design parameters, the method is hence helpful to identify the physical limitations of specific nanostructure models. Note that the full physical response of any point in the reduced dimensionality space can be reconstructed using the decoder part of the autoencoder, also those points that were not present in the training data. 
This means that feasible and non-feasible responses can be analyzed in the original response space (under the assumption that the neural network generalizes well to out-of-training situations).
Note that autoencoders are unsupervised ANN models, which are known to require relatively few data for training. 
This facilitates the application of the technique to experimental data.

In a similar approach, the impact of variations of individual design parameters on the latent space can be studied.
Those parameters whose variations have large (respectively little) impact on the latent space contribute strongly (respectively weakly) to the optical response \cite{kiarashinejadDeepLearningReveals2019, kiarashinejadDeepLearningApproach2019, yeungElucidatingBehaviorNanophotonic2020}.
The latent space is indicated by yellow highlighted neurons in Fig.~\ref{fig:knowledge_discovery}b, top right. 
The impact of physical parameters on these weights is illustrated in the bottom right of figure~\ref{fig:knowledge_discovery}b.
By varying the size of the bottleneck (i.e. reducing the latent space dimension), it is furthermore possible to extract something like the number of principal components of the response, as shown
in the left column of figure~\ref{fig:knowledge_discovery}b.
Iten et al. \cite{itenDiscoveringPhysicalConcepts2020} extended the encoder-decoder ANN for interpretable physics via an approach inspired by humans' interpretation and modelling of physical observations. 
The concept is depicted in Fig.~\ref{fig:knowledge_discovery}c, where the motion of a mass is observed as function of time $x(t)$.
To implement this concept in an ANN the authors append a condition to the latent-vector at the bottleneck of an encoder-decoder ANN (see Fig.~\ref{fig:knowledge_discovery}d). 
This condition is here called a \textit{question}, the example in Fig.~\ref{fig:knowledge_discovery}c uses the time $t'$ for which the ANN shall predict the position of the moving mass ($=$ the \textit{answer}). 
In the context of nano-photonics the question could be an optical spectrum of a nanostructure. 
The ``answer'' returned by the ANN might then be the material or the size of the nanostructure, or a wavelength or laser polarization state. 
This kind of ANN is conceptually very similar to inverse design ANNs (in particular to the cGAN or cAE models), but instead of using it for the design of nanostructures, it is here used to understand causal correlations imposed by the implicit physics in the training data.

A more direct approach to extract physical knowledge from ANNs consists in using the ultra-fast approximation capability of deep learning surrogate models. 
Through a systematic scan of the whole parameter-space it is for example possible to assess the accessible optical responses with a specific nanostructure model. 
In this way, accessible phase and intensity values for metasurface elements have been classified systematically by An et al. \cite{anDeepLearningModeling2020}. 
The logical conclusion of the study was, that allowing more complex shapes for the meta-atoms leads to a larger accessible range for the phase and intensity, as depicted in figure~\ref{fig:knowledge_discovery}e. 
From left to right are shown increasingly complex geometric models (top row) and their accessible scattering phase and intensity range (bottom row).

As already mentioned before, another way to gain insight in physical processes through a machine learning analysis is to use a physical parametrization of the training data, such that the neural network explicitly returns a physical quantity.
As shown in figure~\ref{fig:forward_models}a and~\ref{fig:forward_models}b, extinction spectra can for example be pre-processed in a modal decomposition, such as a superposition of electric and magnetic dipole resonances \cite{soSimultaneousInverseDesign2019} or as a decomposition in Lorentzian resonance profiles \cite{blanchard-dionneTeachingOpticsMachine2020}.
Once trained, the respective neural networks deliver an explicit interpretation of the predicted spectra.

In another recent work, so-called explainable machine learning has been used to assess the importance of constituent parts of a nanostructure with respect to its optical response, as well as to identify those parts of the structure that contribute only weakly to the light-matter interaction \cite{yeungElucidatingDesignBehavior2020}. 
Such information is important for the design of fabrication-robust nanostructures, but also for applications in which sub-constituents of high impact on the nanostructure's optical response need to be identified, e.g. for switchable optical antennas.
Another recent work proposes interpretable machine learning models like decision trees and random forests to understand the physical mechanisms behind inverse design results \cite{elzoukaInterpretableInverseDesign2020}.

\subsection{Deep learning for interpretation of photonics experiments}

The last section of this review is dedicated to recent applications of deep learning in nano-photonics experiments.

Deep learning has proven to enable unprecedented statistical evaluation of large and complicated data, which was formerly impossible with conventional methods.
It has been demonstrated for instance that ANN models can learn from huge microscopy datasets to optically characterize 2D materials such as graphene or transition-metal dichalcogenides (TMDs) \cite{hanDeepLearningEnabled2019} or to automatically localize and classify nano-scale defects \cite{ziatdinovDeepLearningAtomically2017} or to track particles in 3D space using holographic microscopy \cite{shaoMachineLearningHolography2020}.
Deep learning was also successfully applied for the ultra-fast analysis of single molecule emission patterns \cite{zhangAnalyzingComplexSinglemolecule2018} as well as for the experimental reconstruction of quantum states for quantum optics tomography \cite{palmieriExperimentalNeuralNetwork2020}.

By training an ANN on large amounts of experimental optical scattering spectra from complex photonic nanostructures, recently an optical information storage concept has been proposed, able to push the data-density beyond the optical diffraction limit \cite{wiechaPushingLimitsOptical2019}. 
The principle is depicted in figure~\ref{fig:experimental_applications}a. 
Digital information is encoded in silicon nanostructures, which are designed such that each nanostructure encoding a specific bit-sequence possesses a unique scattering spectrum.
Visible light scattering is subsequently interpreted by an artificial neural network (Fig.~\ref{fig:experimental_applications}b). 
Training on experimental data renders this read-out robust against fabrication imperfections and instrumental noise. 
Therein, the ANN is the key ingredient to allow high readout accuracies from distorted data (Fig.~\ref{fig:experimental_applications}c).
Deep learning can be used for various further experimental classification tasks in nano-optics. 
For instance, as depicted in figure~\ref{fig:experimental_applications}d it has been recently demonstrated that an ANN can learn to classify different species of Anthrax spores from holographic microscopy images \cite{joHolographicDeepLearning2017}. 
The confusion rates in the individual classes (Fig.~\ref{fig:experimental_applications}e) allow furthermore to assess similarities and differences between the different Anthrax species.
Similar recent deep-learning based holographic image classification tasks include analysis of colloidal dispersions \cite{yevickMachinelearningApproachHolographic2014} or the real time determination of size and refractive index of subwavelength small particles \cite{midtvedtHolographicCharacterisationSubwavelength2020}.

Deep learning is particularly strong at the interpretation of sparse, undersampled data. 
In a recent example, Argun et al. used a deep neural network for force field calibration in microscopy, by monitoring and interpreting Brownian particle motion \cite{argunEnhancedForcefieldCalibration2020}. 
As depicted in figure~\ref{fig:experimental_applications}f, complex trapping potentials (top left) can be reconstructed efficiently from few experimental samples (top right). 
In contrast to a conventional method (bottom right), the ANN (bottom left) reconstructs the correct potential with high accuracy also from little data (using only the dark part in the top right panel of \ref{fig:experimental_applications}f).
Similarly, machine learning has been used for real-time particle tracking \cite{hannelMachinelearningTechniquesFast2018, newbyConvolutionalNeuralNetworks2018, helgadottirDigitalVideoMicroscopy2019}.
	Recently ANNs have also been succesfully trained on simulated data to efficiently predict the optical forces in complex particle trapping situations \cite{lentonMachineLearningReveals2020}.
Moreover deep learning has been found to be very powerful in solving inverse problems occurring in imaging experiments. 
In this context often sparsity assumptions are required to enable deconstruction of undersampled data, which demands computationally complex inverse solving techniques like compressive sensing.
Corresponding imaging applications include phase recovery \cite{rivensonPhaseRecoveryHolographic2018, nishizakiAnalysisNoniterativePhase2020}, image reconstruction or enhancement \cite{jinDeepConvolutionalNeural2017,  ongieDeepLearningTechniques2020, barbastathisUseDeepLearning2019, rivensonDeepLearningMicroscopy2017}, super-resolution microscopy \cite{nehmeDeepSTORMSuperresolutionSinglemolecule2018, ouyangDeepLearningMassively2018, nehmeDeepSTORM3DDense3D2020, puUnlabeledFarfieldDeeply2020, puLabelfreeDeeplySubwavelength2020} or coherent diffractive imaging \cite{bouchetOptimizingIlluminationPrecise2020, ghoshMachineLearningBased2020}.
In the context of photonics, it has been demonstrated that speckle patterns which occur after light transmission through complex media can be deconstructed very efficiently with deep learning methods \cite{horisakiLearningbasedImagingScattering2016, yunzheDeepSpeckleCorrelation2018, rahmaniMultimodeOpticalFiber2018, borhaniLearningSeeMultimode2018, kurumDeepLearningEnabled2019, bruceFemtometerresolvedSimultaneousMeasurement2020}.
While such speckles appear as if they were random patterns, they are actually the result of deterministic multiple scattering events.
Therefore, a fixed correlation between input and output before and after the complex medium can be established, which is classically done by constructing a transmission matrix \cite{popoffImageTransmissionOpaque2010}, involving complex regularisation schemes, inversion procedures or computationally expensive compressive sensing techniques \cite{frenchSnapshotFiberSpectral2018}. 
While speckle-based methods allow for instance imaging through opaque media or the reconstruction of spectral information, the aforementioned computational burden usually prohibits real time applications.
ANN models on the other hand can be trained to solve the implicit inverse problem in speckle deconstruction very efficiently, which recently enabled to use complex media such as multi-mode fibers for real-time applications in imaging \cite{horisakiLearningbasedImagingScattering2016, yunzheDeepSpeckleCorrelation2018, rahmaniMultimodeOpticalFiber2018, borhaniLearningSeeMultimode2018, pinkardDeepLearningSingleshot2019}, spectral reconstruction \cite{bruceFemtometerresolvedSimultaneousMeasurement2020} or both (hyper-spectral imaging) \cite{kurumDeepLearningEnabled2019}.
Figure~\ref{fig:experimental_applications}g illustrates a setup for such speckle-based hyper-spectral imaging. 
An image is formed via an intensity spatial light modulator (SLM), spectrally shaped using an acousto-optic tunable filter (AOTF), and focused on the aperture of a multi-core multimode fiber bundle (MCMMF). The fiber cores act as pixels of the image, whose individual speckle patterns encode the spectral information. 
K\"ur\"um et al. \cite{kurumDeepLearningEnabled2019} demonstrated that even under noisy conditions and in the undersampling regime, an ANN can reconstruct the spectral information of several thousand fibers with a speed of a few frames per second. 
In contrast, conventional compressive sensing algorithms require tens of minutes for the same task with similar reconstruction fidelity \cite{frenchSnapshotFiberSpectral2018}.

In the context of sparse data reconstruction, deep learning has recently been used in quantum optics applications for the reconstruction of statistical distributions from experiments with weak photon counts, as schematized in figure~\ref{fig:experimental_applications}h.
For instance, Cortes et al. \cite{cortesAcceleratingQuantumOptics2020} demonstrated the successful reconstruction of time dependent data from few photon events using statistical learning. 
In this procedure a machine learning algorithm learns to predict the statistical distribution of the data. 
A similar approach has been applied to assess whether a nano-diamond contains a single or several nitrogen vacancy photon emitters \cite{kudyshevRapidClassificationQuantum2020}. 
An other work demonstrated a machine learning model capable to differentiate between coherent and thermal light sources via a statistical analysis of the temporal distribution of a very low number of photons \cite{youIdentificationLightSources2020}.
These learning based statistical analysis methods are capable to outperform conventional data fitting techniques thanks to their capacity to learn the most probable statistical distributions from the actual data.
Essentially, the machine learning model learns to ``focus'' on the important regions in the data (comparable to adaptive fitting weights). Conventional data fitting algorithms on the other hand bare the risk of attaching too much importance to ``flat'' areas, to the detriment of the accuracy in the relevant regions.
Just as with accidentally over-specialized inverse networks, care must be taken when interpreting the ANN reconstructions. Since data-driven approaches always bare the risk of being biased towards the training data, a neural network might for instance detect a learned statistical distribution even in pure noise.


Deep learning can be applied not only to data analysis, but is also increasingly used to control real time experimental feedback systems. 
Recent examples touching the field of nano-photonics are mainly found in AI-stabilized microscopy.
ANNs can be applied for instance to real-time image enhancement \cite{rivensonDeepLearningEnhanced2018}, microscopy stabilizing feed-back systems \cite{cirovicFeedforwardArtificialNeural1997, liFastConfocalMicroscopy2020} or to conduct sparse data acquisition schemes for the acceleration of scanning microscopy systems via compressive sensing \cite{edePartialScanningTransmission2020}. 
ANNs have been also applied to controlling laser mode-locking stabilization systems \cite{bruntonSelfTuningFiberLasers2014, kutzIntelligentSystemsStabilizing2015, baumeisterDeepLearningModel2018}.
So far the direct application of ANNs to experimental hardware for nano-photonics is still scarce, but the research is in an early stage.
A recent work proposed for instance to calibrate and control electrically reconfigurable photonic circuits by deep learning algorithms \cite{youssryModelingControlReconfigurable2020, youssryModelingControlReconfigurable2020}.
Another example is a pioneering work of Selle et al. \cite{selleModellingUltrafastCoherent2008} which proposed to use ANNs coupled to a femtosecond laser pulse shaper for real-time control of the light-matter interaction in nano-structures or molecules. 
We expect a very rapid development of applications in this direction in the near future, in particular real-time critical applications like sensing  \cite{wangArtificialSensingIntelligence2014} will hugely benefit from the tremendous acceleration potential of ANNs.

\section{Conclusions and perspectives}

In conclusion, in this mini-review we discussed the most recent developments in deep learning methods applied to nano-photonics. 
In the first section we focused on ANN-driven nano-photonic inverse design methods and discussed concepts to improve the design quality of inverse ANNs in comparison with conventional optimization techniques.
In the second part we discussed applications of deep learning in nano-photonics ``beyond inverse design'', spanning from physics informed neural networks over ANNs for physical knowledge extraction to data interpretation and experimental applications.

We would like to emphasize that despite their latest remarkable success and their undeniable great potential, artificial neural networks are ``black boxes''. 
It is extremely hard, mostly even impossible to understand how a neural network generates it's predictions. 
It has been demonstrated at many occasions that even the most sophisticated ANNs, trained on the most carefully assembled datasets contain singular points at which their predictions diverge. 
Another noteworthy danger of data-driven techniques is that they bare a considerable risk to be biased with respect to their training data, like an incident where google's image-tagging algorithm learned implicit racism from its training data \cite{GoogleSaysSorry2015}. 
We therefore appeal to the reader to keep in mind that, simply speaking, ``what you put in is what you get out''. 
In consequence the ANN models are only the second most important ingredient to deep learning. 
The essential element is first of all \textit{the training data}.
Unfortunately it is often understated and not discussed with sufficient emphasis that high quality training data is of utmost importance. By reviewing techniques that aim at improving the training data quality, we tried to arouse some awareness in this respect.
Another important aspect in this context is the \textit{amount} of training data, required to train a well performing and generalizing ANN. 
Unfortunately in many problems which would be naturally suited for deep learning applications, training data is scarce or very expensive to generate. 
Additionally, the more general a problem for an ANN is, the more training data is usually required for a good prediction fidelity.
Last but not least, adapting an ANN model to a new problem often requires the entire training data to be generated from scratch, which might even be the case for minor modifications.
These aspects can create considerable computational barriers for broad and flexible applications of ANNs.

Deep learning techniques in the context of nano-photonics have experienced a tremendous amount of attention in the past few years and research has literally exploded. 
ANNs have enabled manifold applications which formerly seemed strictly impossible. 
As discussed above, a prominent example are data-driven ultra-fast solvers for various inverse problems, for which conventional methods are computationally extremely expensive and slow.
We expect that further groundbreaking applications will be developed in the near future.
For instance very promising progress has been made in the field of Quantum machine learning \cite{schuldIntroductionQuantumMachine2015}, which aims at using deep learning concepts to push the capabilities and interpretability of quantum computing systems. 
In this context, machine learning algorithms recently have autonomously proposed designs for non-trivial quantum optics experiments \cite{krennAutomatedSearchNew2016, melnikovActiveLearningMachine2018, krennComputerinspiredQuantumExperiments2020}.
We expect that deep learning will continue to produce exciting pioneering results. 
We also anticipate that deep learning techniques will become a common numerical tool, regularly employed for the daily use.

	\subsection*{Acknowledgments}
		We thank the NVIDIA Corporation for the donation of a Quadro P6000 GPU used for this research.
		This work was supported by the German Research Foundation (DFG) through a research fellowship (WI 5261/1-1). 
		The authors acknowledge the CALMIP computing facility (grant p20010).
		OM acknowledges support through EPSRC grant EP/M009122/1.

	\subsection*{Disclosures}
	The authors declare no conflicts of interest.


\end{document}